%% file: template_springer.tex
\documentclass[epj2,nopacs]{svjour}
\usepackage{graphics}
\usepackage{epsfig}
\usepackage[pagewise,switch]{lineno}
\def\Journal#1#2#3#4{{#1} {\bf #2} (#4) #3}
\def\NIMA{{ Nucl. Inst. Meth.} \bf A}
\def\etmiss{$E_{\rm T}^{\rm miss}$}

\def\Journal#1#2#3#4{{#1} {\bf #2} (#4) #3}
\def\NIMA{{ Nucl. Inst. Meth.} \bf A}

\begin{document}
\hugehead
\title{Readiness of the ATLAS Liquid Argon Calorimeter for LHC Collisions}
\author{G.~Aad$^{\rm 83}$ \and B.~Abbott$^{\rm 110}$ \and J.~Abdallah$^{\rm
  11}$ \and A.A.~Abdelalim$^{\rm 49}$ \and A.~Abdesselam$^{\rm 117}$
\and O.~Abdinov$^{\rm 10}$ \and B.~Abi$^{\rm 111}$ \and
M.~Abolins$^{\rm 88}$ \and H.~Abramowicz$^{\rm 151}$ \and
H.~Abreu$^{\rm 114}$
% ...
}

\institute{University at Albany, 1400 Washington Ave, Albany, NY
  12222, United States of America \and University of Alberta,
  Department of Physics, Centre for Particle Physics, Edmonton, AB T6G
  2G7, Canada \and Ankara University, Faculty of Sciences, Department
  of Physics, TR 061000 Tandogan, Ankara, Turkey \and LAPP,
  Universit\'e de Savoie, CNRS/IN2P3, Annecy-le-Vieux, France \and
  Argonne National Laboratory, High Energy Physics Division, 9700
  S. Cass Avenue, Argonne IL 60439, United States of America \and
  University of Arizona, Department of Physics, Tucson, AZ 85721,
  United States of America \and The University of Texas at Arlington,
  Department of Physics, Box 19059, Arlington, TX 76019, United States
  of America \and University of Athens, Nuclear \& Particle Physics,
  Department of Physics, Panepistimiopouli, Zografou, GR 15771 Athens,
  Greece \and National Technical University of Athens, Physics
  Department, 9-Iroon Polytechniou, GR 15780 Zografou, Greece \and
  Institute of Physics, Azerbaijan Academy of Sciences, H. Javid
  Avenue 33, AZ 143 Baku, Azerbaijan \and Institut de F\'isica d'Altes
  Energies, IFAE, Edifici Cn, Universitat Aut\`onoma  de Barcelona,
  ES - 08193 Bellaterra (Barcelona), Spain \and $^{(a)}$University of
  Belgrade, Institute of Physics, P.O. Box 57, 11001 Belgrade; Vinca
  Institute of Nuclear Sciences$^{(b)}$, Mihajla Petrovica Alasa
  12-14, 11001 Belgrade, Serbia \and University of Bergen, Department
  for Physics and Technology, Allegaten 55, NO - 5007 Bergen, Norway
  \and Lawrence Berkeley National Laboratory and University of
  California, Physics Division, MS50B-6227, 1 Cyclotron Road,
  Berkeley, CA 94720, United States of America \and Humboldt
  University, Institute of Physics, Berlin, Newtonstr. 15, D-12489
  Berlin, Germany \and University of Bern \and Albert Einstein Center
  for Fundamental Physics,
 Laboratory for High Energy Physics,
  Sidlerstrasse 5, CH - 3012 Bern, Switzerland\\ University of
  Birmingham, School of Physics and Astronomy, Edgbaston, Birmingham
  B15 2TT, United Kingdom \and Bogazici University, Faculty of
  Sciences, Department of Physics, TR - 80815 Bebek-Istanbul, Turkey
  \and INFN Sezione di Bologna$^{(a)}$; Universit\`a  di Bologna,
  Dipartimento di Fisica$^{(b)}$, viale C. Berti Pichat, 6/2, IT -
  40127 Bologna, Italy \and University of Bonn, Physikalisches
  Institut, Nussallee 12, D - 53115 Bonn, Germany \and Boston
  University, Department of Physics,  590 Commonwealth Avenue, Boston,
  MA 02215, United States of America \and Brandeis University,
  Department of Physics, MS057, 415 South Street, Waltham, MA 02454,
  United States of America \and Universidade Federal do Rio De
  Janeiro, Instituto de Fisica$^{(a)}$, Caixa Postal 68528, Ilha do
  Fundao, BR - 21945-970 Rio de Janeiro; $^{(b)}$Universidade de Sao
  Paulo, Instituto de Fisica, R.do Matao Trav. R.187, Sao Paulo - SP,
  05508 - 900, Brazil \and Brookhaven National Laboratory, Physics
  Department, Bldg. 510A, Upton, NY 11973, United States of America
  \and National Institute of Physics and Nuclear Engineering$^{(a)}$,
  Bucharest-Magurele, Str. Atomistilor 407,  P.O. Box MG-6, R-077125,
  Romania; $^{(b)}$University Politehnica Bucharest, Rectorat - AN
  001, 313 Splaiul Independentei, sector 6, 060042 Bucuresti;
  $^{(c)}$West University in Timisoara, Bd. Vasile Parvan 4,
  Timisoara, Romania \and Universidad de Buenos Aires, FCEyN,
  Dto. Fisica, Pab I - C. Universitaria, 1428 Buenos Aires, Argentina
  \and University of Cambridge, Cavendish Laboratory, J J Thomson
  Avenue, Cambridge CB3 0HE, United Kingdom \and Carleton University,
  Department of Physics, 1125 Colonel By Drive, Ottawa ON  K1S 5B6,
  Canada \and CERN, CH - 1211 Geneva 23, Switzerland \and University
  of Chicago, Enrico Fermi Institute, 5640 S. Ellis Avenue, Chicago,
  IL 60637, United States of America \and Pontificia Universidad
  Cat\'olica de Chile, Facultad de Fisica, Departamento de
  Fisica$^{(a)}$, Avda. Vicuna Mackenna 4860, San Joaquin, Santiago;
  Universidad T\'ecnica Federico Santa Mar\'ia, Departamento de
  F\'isica$^{(b)}$, Avda. Esp\~ana 1680, Casilla 110-V, Valpara\'iso,
  Chile \and Institute of HEP, Chinese Academy of Sciences, P.O. Box
  918, CN-100049 Beijing; USTC, Department of Modern Physics, Hefei,
  CN-230026 Anhui; Nanjing University, Department of Physics,
  CN-210093 Nanjing; Shandong University, HEP Group, CN-250100
  Shadong, China \and Laboratoire de Physique Corpusculaire,
  CNRS-IN2P3, Universit\'e Blaise Pascal, FR - 63177 Aubiere Cedex,
  France \and Columbia University, Nevis Laboratory, 136 So. Broadway,
  Irvington, NY 10533, United States of America \and University of
  Copenhagen, Niels Bohr Institute, Blegdamsvej 17, DK - 2100
  Kobenhavn 0, Denmark \and INFN Gruppo Collegato di Cosenza$^{(a)}$;
  Universit\`a della Calabria, Dipartimento di Fisica$^{(b)}$,
  IT-87036 Arcavacata di Rende, Italy \and Faculty of Physics and
  Applied Computer Science of the AGH-University of Science and
  Technology, (FPACS, AGH-UST), al. Mickiewicza 30, PL-30059 Cracow,
  Poland \and The Henryk Niewodniczanski Institute of Nuclear Physics,
  Polish Academy of Sciences, ul. Radzikowskiego 152, PL - 31342
  Krakow, Poland \and Southern Methodist University, Physics
  Department, 106 Fondren Science Building, Dallas, TX 75275-0175,
  United States of America \and University of Texas at Dallas, 800
  West Campbell Road, Richardson, TX 75080-3021, United States of
  America \and DESY, Notkestr. 85, D-22603 Hamburg , Germany and
  Platanenalle 6, D-15738 Zeuthen, Germany \and TU Dortmund,
  Experimentelle Physik IV, DE - 44221 Dortmund, Germany \and
  Technical University Dresden, Institut fuer Kern- und
  Teilchenphysik, Zellescher Weg 19, D-01069 Dresden, Germany \and
  Duke University, Department of Physics, Durham, NC 27708, United
  States of America \and University of Edinburgh, School of Physics \&
  Astronomy, James Clerk Maxwell Building, The Kings Buildings,
  Mayfield Road, Edinburgh EH9 3JZ, United Kingdom \and Fachhochschule
  Wiener Neustadt; Johannes Gutenbergstrasse 3 AT - 2700 Wiener
  Neustadt, Austria \and INFN Laboratori Nazionali di Frascati, via
  Enrico Fermi 40, IT-00044 Frascati, Italy \and
  Albert-Ludwigs-Universit\"{a}t, Fakult\"{a}t f\"{u}r Mathematik und
  Physik, Hermann-Herder Str. 3, D - 79104 Freiburg i.Br., Germany
  \and Universit\'e de Gen\`eve, Section de Physique, 24 rue Ernest
  Ansermet, CH - 1211 Geneve 4, Switzerland \and INFN Sezione di
  Genova$^{(a)}$; Universit\`a  di Genova, Dipartimento di
  Fisica$^{(b)}$, via Dodecaneso 33, IT - 16146 Genova, Italy \and
  Institute of Physics of the Georgian Academy of Sciences, 6
  Tamarashvili St., GE - 380077 Tbilisi; Tbilisi State University, HEP
  Institute, University St. 9, GE - 380086 Tbilisi, Georgia \and
  Justus-Liebig-Universitaet Giessen, II Physikalisches Institut,
  Heinrich-Buff Ring 16,  D-35392 Giessen, Germany \and University of
  Glasgow, Department of Physics and Astronomy, Glasgow G12 8QQ,
  United Kingdom \and Georg-August-Universitat, II. Physikalisches
  Institut, Friedrich-Hund Platz 1, D-37077 Goettingen, Germany \and
  Laboratoire de Physique Subatomique et de Cosmologie, CNRS/IN2P3,
  Universit\'e Joseph Fourier, INPG, 53 avenue des Martyrs, FR - 38026
  Grenoble Cedex, France \and Hampton University, Department of
  Physics, Hampton, VA 23668, United States of America \and Harvard
  University, Laboratory for Particle Physics and Cosmology, 18
  Hammond Street, Cambridge, MA 02138, United States of America \and
  Ruprecht-Karls-Universitaet Heidelberg, Kirchhoff-Institut fuer
  Physik$^{(a)}$, Im Neuenheimer Feld 227, DE - 69120 Heidelberg;
  $^{(b)}$Physikalisches Institut, Philosophenweg 12, D-69120
  Heidelberg; ZITI Ruprecht-Karls-University Heidelberg$^{(c)}$,
  Lehrstuhl fuer Informatik V, B6, 23-29, DE - 68131 Mannheim, Germany
  \and Hiroshima University, Faculty of Science, 1-3-1 Kagamiyama,
  Higashihiroshima-shi, JP - Hiroshima 739-8526, Japan \and Hiroshima
  Institute of Technology, Faculty of Applied Information Science,
  2-1-1 Miyake Saeki-ku, Hiroshima-shi, JP - Hiroshima 731-5193, Japan
  \and Indiana University, Department of Physics,  Swain Hall West
  117, Bloomington, IN 47405-7105, United States of America \and
  Institut fuer Astro- und Teilchenphysik, Technikerstrasse 25, A -
  6020 Innsbruck, Austria \and University of Iowa, 203 Van Allen Hall,
  Iowa City, IA 52242-1479, United States of America \and Iowa State
  University, Department of Physics and Astronomy, Ames High Energy
  Physics Group,  Ames, IA 50011-3160, United States of America \and
  Joint Institute for Nuclear Research, JINR Dubna, RU - 141 980
  Moscow Region, Russia \and KEK, High Energy Accelerator Research
  Organization, 1-1 Oho, Tsukuba-shi, Ibaraki-ken 305-0801, Japan \and
  Kobe University, Graduate School of Science, 1-1 Rokkodai-cho,
  Nada-ku, JP Kobe 657-8501, Japan \and Kyoto University, Faculty of
  Science, Oiwake-cho, Kitashirakawa, Sakyou-ku, Kyoto-shi, JP - Kyoto
  606-8502, Japan \and Kyoto University of Education, 1 Fukakusa,
  Fujimori, fushimi-ku, Kyoto-shi, JP - Kyoto 612-8522, Japan \and
  Universidad Nacional de La Plata, FCE, Departamento de F\'{i}sica,
  IFLP (CONICET-UNLP),   C.C. 67,  1900 La Plata, Argentina \and
  Lancaster University, Physics Department, Lancaster LA1 4YB, United
  Kingdom \and INFN Sezione di Lecce$^{(a)}$; Universit\`a  del
  Salento, Dipartimento di Fisica$^{(b)}$Via Arnesano IT - 73100
  Lecce, Italy \and University of Liverpool, Oliver Lodge Laboratory,
  P.O. Box 147, Oxford Street,  Liverpool L69 3BX, United Kingdom \and
  Jo\v{z}ef Stefan Institute and University of Ljubljana, Department
  of Physics, SI-1000 Ljubljana, Slovenia \and Queen Mary University
  of London, Department of Physics, Mile End Road, London E1 4NS,
  United Kingdom \and Royal Holloway, University of London, Department
  of Physics, Egham Hill, Egham, Surrey TW20 0EX, United Kingdom \and
  University College London, Department of Physics and Astronomy,
  Gower Street, London WC1E 6BT, United Kingdom \and Laboratoire de
  Physique Nucl\'eaire et de Hautes Energies, Universit\'e Pierre et
  Marie Curie (Paris 6), Universit\'e Denis Diderot (Paris-7),
  CNRS/IN2P3, Tour 33, 4 place Jussieu, FR - 75252 Paris Cedex 05,
  France \and Lunds universitet, Naturvetenskapliga fakulteten,
  Fysiska institutionen, Box 118, SE - 221 00 Lund, Sweden \and
  Universidad Autonoma de Madrid, Facultad de Ciencias, Departamento
  de Fisica Teorica, ES - 28049 Madrid, Spain \and Universitaet Mainz,
  Institut fuer Physik, Staudinger Weg 7, DE - 55099 Mainz, Germany
  \and University of Manchester, School of Physics and Astronomy,
  Manchester M13 9PL, United Kingdom \and CPPM, Aix-Marseille
  Universit\'e, CNRS/IN2P3, Marseille, France \and University of
  Massachusetts, Department of Physics, 710 North Pleasant Street,
  Amherst, MA 01003, United States of America \and McGill University,
  High Energy Physics Group, 3600 University Street, Montreal, Quebec
  H3A 2T8, Canada \and University of Melbourne, School of Physics, AU
  - Parkville, Victoria 3010, Australia \and The University of
  Michigan, Department of Physics, 2477 Randall Laboratory, 500 East
  University, Ann Arbor, MI 48109-1120, United States of America \and
  Michigan State University, Department of Physics and Astronomy, High
  Energy Physics Group, East Lansing, MI 48824-2320, United States of
  America \and INFN Sezione di Milano$^{(a)}$; Universit\`a  di
  Milano, Dipartimento di Fisica$^{(b)}$, via Celoria 16, IT - 20133
  Milano, Italy \and B.I. Stepanov Institute of Physics, National
  Academy of Sciences of Belarus, Independence Avenue 68, Minsk
  220072, Republic of Belarus \and National Scientific \& Educational
  Centre for Particle \& High Energy Physics, NC PHEP BSU,
  M. Bogdanovich St. 153, Minsk 220040, Republic of Belarus \and
  Massachusetts Institute of Technology, Department of Physics, Room
  24-516, Cambridge, MA 02139, United States of America \and
  University of Montreal, Group of Particle Physics, C.P. 6128,
  Succursale Centre-Ville, Montreal, Quebec, H3C 3J7  , Canada \and
  P.N. Lebedev Institute of Physics, Academy of Sciences, Leninsky
  pr. 53, RU - 117 924 Moscow, Russia \and Institute for Theoretical
  and Experimental Physics (ITEP), B. Cheremushkinskaya ul. 25, RU 117
  218 Moscow, Russia \and Moscow Engineering \& Physics Institute
  (MEPhI), Kashirskoe Shosse 31, RU - 115409 Moscow, Russia \and
  Lomonosov Moscow State University Skobeltsyn Institute of Nuclear
  Physics (MSU SINP), 1(2), Leninskie gory, GSP-1, Moscow 119991
  Russian Federation , Russia \and Ludwig-Maximilians-Universit\"at
  M\"unchen, Fakult\"at f\"ur Physik, Am Coulombwall 1,  DE - 85748
  Garching, Germany \and Max-Planck-Institut f\"ur Physik,
  (Werner-Heisenberg-Institut), F\"ohringer Ring 6, 80805 M\"unchen,
  Germany \and Nagasaki Institute of Applied Science, 536 Aba-machi,
  JP Nagasaki 851-0193, Japan \and Nagoya University, Graduate School
  of Science, Furo-Cho, Chikusa-ku, Nagoya, 464-8602, Japan \and INFN
  Sezione di Napoli$^{(a)}$; Universit\`a  di Napoli, Dipartimento di
  Scienze Fisiche$^{(b)}$, Complesso Universitario di Monte
  Sant'Angelo, via Cinthia, IT - 80126 Napoli, Italy \and  University
  of New Mexico, Department of Physics and Astronomy, MSC07  4220,
  Albuquerque, NM 87131 USA, United States of America \and Radboud
  University Nijmegen/NIKHEF, Department of Experimental High Energy
  Physics, Toernooiveld 1, NL - 6525 ED Nijmegen , Netherlands \and
  Nikhef National Institute for Subatomic Physics, and University of
  Amsterdam, Science Park 105, 1098 XG Amsterdam, Netherlands \and
  Budker Institute of Nuclear Physics (BINP), RU - Novosibirsk 630
  090, Russia \and New York University, Department of Physics, 4
  Washington Place, New York NY 10003, USA, United States of America
  \and Ohio State University, 191 West Woodruff Ave, Columbus, OH
  43210-1117, United States of America \and Okayama University,
  Faculty of Science, Tsushimanaka 3-1-1, Okayama 700-8530, Japan \and
  University of Oklahoma, Homer L. Dodge Department of Physics and
  Astronomy, 440 West Brooks, Room 100, Norman, OK 73019-0225, United
  States of America \and Oklahoma State University, Department of
  Physics, 145 Physical Sciences Building, Stillwater, OK 74078-3072,
  United States of America \and Palack\'y University, 17.listopadu
  50a,  772 07  Olomouc, Czech Republic \and University of Oregon,
  Center for High Energy Physics, Eugene, OR 97403-1274, United States
  of America \and LAL, Univ. Paris-Sud, IN2P3/CNRS, Orsay, France \and
  Osaka University, Graduate School of Science, Machikaneyama-machi
  1-1, Toyonaka, Osaka 560-0043, Japan \and University of Oslo,
  Department of Physics, P.O. Box 1048,  Blindern, NO - 0316 Oslo 3,
  Norway \and Oxford University, Department of Physics, Denys
  Wilkinson Building, Keble Road, Oxford OX1 3RH, United Kingdom \and
  INFN Sezione di Pavia$^{(a)}$; Universit\`a  di Pavia, Dipartimento
  di Fisica Nucleare e Teorica$^{(b)}$, Via Bassi 6, IT-27100 Pavia,
  Italy \and University of Pennsylvania, Department of Physics, High
  Energy Physics Group, 209 S. 33rd Street, Philadelphia, PA 19104,
  United States of America \and Petersburg Nuclear Physics Institute,
  RU - 188 300 Gatchina, Russia \and INFN Sezione di Pisa$^{(a)}$;
  Universit\`a   di Pisa, Dipartimento di Fisica E. Fermi$^{(b)}$,
  Largo B. Pontecorvo 3, IT - 56127 Pisa, Italy \and University of
  Pittsburgh, Department of Physics and Astronomy, 3941 O'Hara Street,
  Pittsburgh, PA 15260, United States of America \and
  $^{(a)}$Universidad de Granada, Departamento de Fisica Teorica y del
  Cosmos and CAFPE, E-18071 Granada; Laboratorio de Instrumentacao e
  Fisica Experimental de Particulas - LIP$^{(b)}$, Avenida Elias
  Garcia 14-1, PT - 1000-149 Lisboa, Portugal \and Institute of
  Physics, Academy of Sciences of the Czech Republic, Na Slovance 2,
  CZ - 18221 Praha 8, Czech Republic \and Charles University in
  Prague, Faculty of Mathematics and Physics, Institute of Particle
  and Nuclear Physics, V Holesovickach 2, CZ - 18000 Praha 8, Czech
  Republic \and Czech Technical University in Prague, Zikova 4, CZ -
  166 35 Praha 6, Czech Republic \and State Research Center Institute
  for High Energy Physics, Moscow Region, 142281, Protvino, Pobeda
  street, 1, Russia \and Rutherford Appleton Laboratory, Science and
  Technology Facilities Council, Harwell Science and Innovation
  Campus, Didcot OX11 0QX, United Kingdom \and University of Regina,
  Physics Department, Canada \and Ritsumeikan University, Noji Higashi
  1 chome 1-1, JP - Kusatsu, Shiga 525-8577, Japan \and INFN Sezione
  di Roma I$^{(a)}$; Universit\`a  La Sapienza, Dipartimento di
  Fisica$^{(b)}$, Piazzale A. Moro 2, IT- 00185 Roma, Italy \and INFN
  Sezione di Roma Tor Vergata$^{(a)}$; Universit\`a di Roma Tor
  Vergata, Dipartimento di Fisica$^{(b)}$ , via della Ricerca
  Scientifica, IT-00133 Roma, Italy \and INFN Sezione di  Roma
  Tre$^{(a)}$; Universit\`a Roma Tre, Dipartimento di Fisica$^{(b)}$,
  via della Vasca Navale 84, IT-00146  Roma, Italy \and Universit\'e
  Hassan II, Facult\'e des Sciences Ain Chock$^{(a)}$, B.P. 5366, MA -
  Casablanca; Centre National de l'Energie des Sciences Techniques
  Nucleaires (CNESTEN)$^{(b)}$, B.P. 1382 R.P. 10001 Rabat 10001;
  Universit\'e Mohamed Premier$^{(c)}$, LPTPM, Facult\'e des Sciences,
  B.P.717. Bd. Mohamed VI, 60000, Oujda ; Universit\'e Mohammed V,
  Facult\'e des Sciences$^{(d)}$, LPNR, BP 1014, 10000 Rabat, Morocco
  \and CEA, DSM/IRFU, Centre d'Etudes de Saclay, FR - 91191
  Gif-sur-Yvette, France \and University of California Santa Cruz,
  Santa Cruz Institute for Particle Physics (SCIPP), Santa Cruz, CA
  95064, United States of America \and University of Washington,
  Seattle, Department of Physics, Box 351560, Seattle, WA 98195-1560,
  United States of America \and University of Sheffield, Department of
  Physics \& Astronomy, Hounsfield Road, Sheffield S3 7RH, United
  Kingdom \and Shinshu University, Department of Physics, Faculty of
  Science, 3-1-1 Asahi, Matsumoto-shi, JP - Nagano 390-8621, Japan
  \and Universitaet Siegen, Fachbereich Physik, D 57068 Siegen,
  Germany \and Simon Fraser University, Department of Physics, 8888
  University Drive, CA - Burnaby, BC V5A 1S6, Canada \and SLAC
  National Accelerator Laboratory, Stanford, California 94309, United
  States of America \and Comenius University, Faculty of Mathematics,
  Physics \& Informatics, Mlynska dolina F2, SK - 84248 Bratislava;
  Institute of Experimental Physics of the Slovak Academy of Sciences,
  Dept. of Subnuclear Physics, Watsonova 47, SK - 04353 Kosice, Slovak
  Republic \and Stockholm University, Department of Physics, AlbaNova,
  SE - 106 91 Stockholm, Sweden \and Royal Institute of Technology
  (KTH), Physics Department, SE - 106 91 Stockholm, Sweden \and Stony
  Brook University, Department of Physics and Astronomy, Nicolls Road,
  Stony Brook, NY 11794-3800, United States of America \and University
  of Sussex, Department of Physics and Astronom \and Pevensey 2
  Building, Falmer, Brighton BN1 9QH, United Kingdom\\ University of
  Sydney, School of Physics, AU - Sydney NSW 2006, Australia \and
  Insitute of Physics, Academia Sinica, TW - Taipei 11529, Taiwan \and
  Technion, Israel Inst. of Technology, Department of Physics,
  Technion City, IL - Haifa 32000, Israel \and Tel Aviv University,
  Raymond and Beverly Sackler School of Physics and Astronomy, Ramat
  Aviv, IL - Tel Aviv 69978, Israel
% ...
}
\date{Received: date / Revised version: date}
\abstract{
The ATLAS liquid argon calorimeter has been operating continuously
since August 2006. At this time, only part of the calorimeter was
readout, but since the beginning of 2008, all calorimeter cells have been
connected to the ATLAS readout system in preparation for LHC
collisions. This paper gives an overview of the liquid argon
calorimeter performance measured in situ with random triggers,
calibration data, cosmic muons, and LHC beam splash events. Results on
the detector operation, timing performance, electronics noise, and
gain stability are presented. High energy deposits from radiative
cosmic muons and beam splash events allow to check the
intrinsic constant term of the energy resolution. The uniformity of the
electromagnetic
barrel calorimeter response along $\eta$ (averaged over $\phi$) is
measured at the percent level using minimum ionizing cosmic
muons. Finally, studies of electromagnetic showers from radiative
muons have been used to cross-check the Monte Carlo simulation. The
performance results obtained using the ATLAS readout, data
acquisition, and reconstruction software indicate that the liquid
argon calorimeter is well-prepared for collisions at the dawn of the
LHC era.
} 
\maketitle
%

\runningpagewiselinenumbers

\input{Introduction}

\input{CaloStatus_epj}

\input{NoisePedestal_epj}

\input{Timing_epj}

\input{ErecQuality_epj}

\input{Uniformity_epj}
\input{Photons_epj}

\input{Conclusion}
\input{Acknowledgement-25may09}

\input{Biblio}

\end{document}

%% file: Introduction.tex
\section{Introduction}
\label{sec:intro}

Installation of the liquid argon (LAr) calorimeter in the
ATLAS~\cite{ATLAS_PHYS_TDR} experimental hall was completed in early
2008. Until recently, the expected performance of the LAr calorimeter
was extrapolated from intensive testing of a few modules with electron
and pion beams from 1998 to 2003
(Ref.~\cite{ATLAS_TB_EM1,ATLAS_TB_EM2,ATLAS_TB_EM3,ATLAS_TB_EM4,ATLAS_TB_EM5,ATLAS_TB_HEC1,ATLAS_TB_HEC2,ATLAS_TB_FCAL1,ATLAS_TB_FCAL2}), and in 2004 of a
complete ATLAS detector 
slice~\cite{ATLAS_CTB1,ATLAS_CTB2,ATLAS_CTB3}. The 20 months
separating the completion of the installation from the first LHC
collisions have been used to commission the LAr calorimeter. This
paper reviews the first in situ measurements of the
electronics stability, the quality of the energy reconstruction, 
the calorimeter response uniformity and the agreement between data
and the Monte Carlo simulation of electromagnetic shower shapes. The
measurements are performed using calibration triggers, cosmic muons,
and the first LHC beam events collected during this 20 months period. The
results and the experience gained in the operation of the LAr
calorimeter provide the foundation for a more rapid understanding of
the experimental signatures of the first LHC collisions, involving
electrons, photons, missing transverse energy (\etmiss), jets, and $\tau$s
where the LAr calorimeter plays a central role. 

This paper is organized as follows. Section~\ref{sec:CaloStatus} gives
the present hardware status of the LAr
calorimeter. Section~\ref{sec:Erec} details the level of understanding
of the ingredients entering the cell energy reconstruction: pedestals,
noise, electronic gains, timing, and the quality of the signal pulse
shape predictions. The current understanding of the first level
trigger energy computation is also
discussed. Section~\ref{sec:analysis} describes the in situ
performance of the electromagnetic LAr calorimeter using ionizing and
radiating cosmic muons. Lastly, Section~\ref{sec:conclu} draws the
conclusions.

%% file: CaloStatus_epj.tex
\section{LAr calorimeter hardware status and data taking conditions}
\label{sec:CaloStatus}

The LAr calorimeter is composed of electromagnetic and hadronic
sub-detectors of which the main characteristics are described in
Section~\ref{sec:CaloDescr}. During the detector and electronics
construction and installation, regular and stringent quality tests
were performed, resulting in a fully functional LAr calorimeter. The
operational stability of the cryostats since March 2008 is discussed
in Section~\ref{sec:CryoOp}. The current status of the high voltage
and  the cell readout are discussed in Sections~\ref{sec:CaloHV}
and~\ref{sec:CaloReadOut} respectively. Finally, the general data
taking conditions are given in Section~\ref{sec:Data}. In ATLAS, the
positive $x$-axis is defined as pointing from the interaction point to
the center of the LHC ring, the positive $y$-axis is defined as
pointing upwards, and the positive $z$-axis corresponds to protons
running anti-clockwise. The polar angle $\theta$ is measured from the
beam axis ($z$-axis), the azimuthal angle $\phi$ is measured in the
transverse ($xy$)-plane, and the pseudorapidity is defined as $\eta =$
--ln tan($\theta$/2). 

\subsection{Main characteristics of the LAr calorimeter}
\label{sec:CaloDescr}

\begin{figure}[hbtp]
\begin{center}
\includegraphics[width=0.45\textwidth]{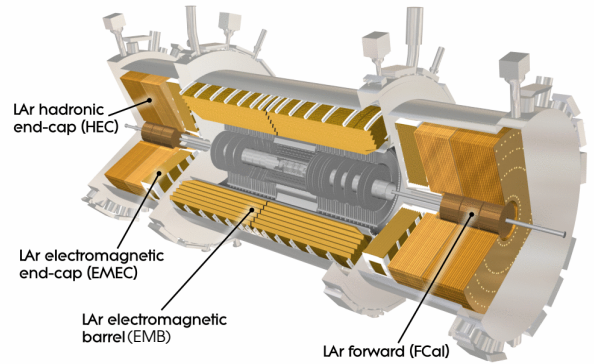}
\end{center}
\caption{\it Cut-away view of the LAr calorimeter, 17 m long (barrel +
  endcaps) and 4 m of diameter.}
\label{fig:LArCalo}
\end{figure}

The LAr calorimeter~\cite{ATLAS_PHYS_TDR}, shown in Figure~\ref{fig:LArCalo}, is 
composed of sampling detectors with full azimuthal symmetry,
housed in one barrel and two endcap cryostats. More
specifically, a highly granular electromagnetic (EM) calorimeter with
accordion-shaped electrodes and lead absorbers in liquid argon covers the
pseudorapidity range $|\eta| < 3.2$, and contains a barrel part 
(EMB~\cite{EMB_CONS}, $|\eta|<1.475$) and an endcap part 
(EMEC~\cite{EMEC_CONS}, $1.375<|\eta|<3.2$). For $|\eta|< 1.8$, 
a presampler (PS~\cite{PS_CONS,EMEC_CONS}), consisting of an active LAr
layer and installed directly in front of 
the EM calorimeters, provides a measurement of the energy lost 
upstream. Located behind the EMEC is a copper-liquid argon hadronic
endcap calorimeter (HEC~\cite{HEC_CONS}, $1.5<|\eta|<3.2$), and a 
copper/tungsten-liquid argon forward calorimeter (FCal \cite{FCAL_CONS}) 
covers the region closest to the beam at $3.1<|\eta|<4.9$. An hadronic
Tile calorimeter ($|\eta|<1.7$) surrounding the LAr cryostats completes
the ATLAS calorimetry.

All the LAr detectors are segmented transversally and divided in three or four layers in 
depth, and correspond to a total of 182,468 readout cells, i.e. 97.2\% of the full
ATLAS calorimeter readout. 

The relative energy resolution of the LAr calorimeter is usually parameterized by:
\begin{equation}
\frac{\sigma_E}{E}=\frac{a}{\sqrt{E}} \oplus \frac{b}{E} \oplus c,
\label{eq:Eresolution}
\end{equation}
where ($a$) is the stochastic term, ($b$) the noise term and ($c$) the
constant term. The target values for these terms are respectively
$a\simeq 10\%$, $b\simeq 170$ MeV (without pile-up) and $c=0.7\%$. 

\subsection{Cryostat operation}
\label{sec:CryoOp}

\begin{figure}[hbtp]
\begin{center}
\includegraphics[width=0.45\textwidth]{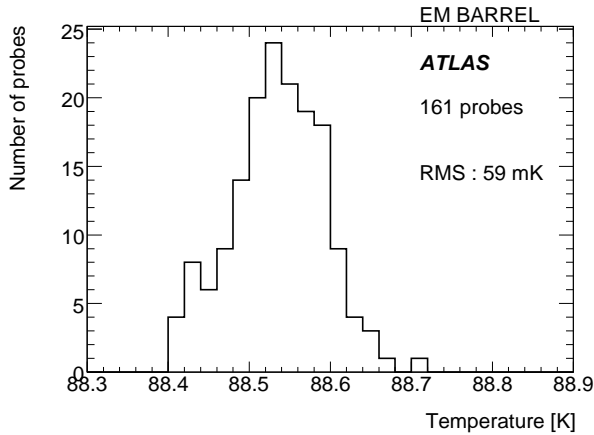}
\end{center}
\caption{Distribution of barrel cryostat probe temperatures averaged over a period of ten days.}
\label{fig:TempPur}
\end{figure}

Variations of the liquid argon temperature have a direct impact 
on the readout signal, and consequently on the energy scale, partly 
through the effect on the argon density, but mostly through
the effect on the ionization electron drift
velocity in the LAr. Overall, a $-2\%$/K signal variation
is expected~\cite{TDRIFT}. The need to keep the corresponding 
contribution to the constant term of the energy resolution (Eq.~\ref{eq:Eresolution})
negligible (i.e. well below 0.2\%)
imposes a temperature uniformity requirement of better than 
100 mK in each cryostat. In the liquid, $\sim$500 
temperature probes (PT100 platinum resistors) 
are fixed on the LAr detector components 
and read out every minute. In 2008-2009, installation activities in
the ATLAS cavern prevented a stable cryostat temperature. A quiet 
period of ten days around the 2008 Christmas break, representative
of what is expected during LHC collisions, allowed a check of the
temperature stability in the absence of these external factors. 
The average dispersion (RMS) of the measurements of each
temperature probe over this period is 1.6 mK (5 mK maximum), showing that no significant local
temperature variation in time is observed in the three cryostats. Over this period, the
temperature uniformity (RMS of all probes per cyostat) is illustrated for the 
barrel in Figure~\ref{fig:TempPur} and gives 59 mK. Results for the two endcap cryostats
are also in the range 50-70 mK, below the required level of 100~mK. The average cryostat
temperatures are slightly different for the barrel (88.49 K) and the
two endcaps (88.67 and 88.45 K) because they are independently
regulated. An energy scale correction per cryostat will therefore be
applied. 

To measure the effects of possible out-gassing of calorimeter materials under irradiation, 
which has been minimized by careful screening of components, 30 purity monitors 
measuring the energy deposition of radioactive sources in the LAr are installed 
in each cryostat and read every 15 minutes. 
The contribution to the constant term of the energy resolution is negligible for a level 
of electronegative impurities below 1000~ppb $O_2$ equivalent. All argon 
purity measurements over a period of two years are stable, in the
range $200\pm100$ ppb $O_2$ equivalent, well below this requirement. 

In summary, measurements of the liquid argon temperature and purity
demonstrate that the stability of the operation of the three LAr
cryostats is in the absence of proton beams within the required 
limits ensuring a
negligible contribution to the energy resolution constant term.

\subsection{High voltage status}
\label{sec:CaloHV}

The electron/ion drift speed in the LAr gap depends on the electric
field, typically 1~kV/mm. Sub-detector-specific high voltage (HV)
settings are applied.
In the EM barrel, the high voltage is constant along $\eta$, while in
the EMEC, where the gap varies continuously with radius, it is adjusted  
in steps along $\eta$. The HV supply granularity is typically in sectors of
$\Delta\eta \times \Delta\phi = 0.2\times0.2$. For
redundancy, each side of an EM electrode, which is in the middle of the LAr gap, 
is powered separately. In the HEC, each sub-gap is serviced by one of
four different HV lines, while for the FCal each of the four electrode
groups forming a normal readout channel is served by an independent HV line. 

For HV sectors with non-optimal behavior, solutions were implemented in
order to recover the corresponding region. For example, in the EM calorimeter, 
faulty electrodes were connected to separate HV lines during the 
assembly phase at room temperature while, if the defect was identified 
during cryostat cold testing, the high voltage sector was divided into
two in $\phi$, each connected separately. The effect of zero voltage 
on one side of an electrode was studied
in beam tests proving that with offline corrections
the energy can still be measured, with only a small loss in accuracy. Finally, for HV sectors
with a permanent short-circuit, high voltage modules permitting large DC
current draws of up to 3 mA (more than three orders of magnitude above the nominal limit) 
are used in order to operate the faulty sector at 1000 V or above.

\begin{figure}[hbtp]
\begin{center}
\includegraphics[width=0.45\textwidth]{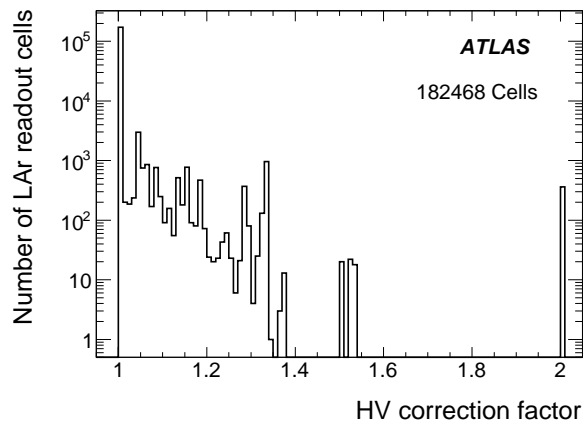}
\end{center}
\caption{High voltage correction factors for all LAr cells at the end of September 2009.}
\label{fig:HVCor}
\end{figure}

As a result, $93.9\%$ of readout cells are operating under nominal
conditions and the rest sees a reduced high voltage. However, even
with a reduced high voltage, signals can be well reconstructed by using
a correction scale factor. Figure~\ref{fig:HVCor} shows 
the distribution of all HV correction factors for the EM, HEC and FCal cells as of the
end of September 2009. Since the beginning of 2008, no changes have been observed.
The largest correction occurs if one side of an EM electrode is not powered, and
only half of the signal is collected. For the faulty cells, this correction 
factor is applied online at the energy 
reconstruction level. A similar correction is currently being implemented at
the first level (L1) trigger.   

In conclusion, since the beginning of 2008, all 182,468 readout cells
are powered with high voltage, and no dead region exists. Signals 
from regions with non-nominal high voltage are easily corrected and their impact on 
physics is negligible.

\subsection{Readout cell status}
\label{sec:CaloReadOut}

The cell signals are read out through 1524 Front-End Boards (FEBs~\cite{FEB_CONS,ALLFEB_CONS}) with 128
channels each, which sit inside front-end crates that are located
around the periphery of the cryostats. The FEBs perform
analog processing (amplification and shaping - except for the HEC
where the amplification is done inside the cryostat), store the signal while
waiting for the L1 trigger decision, and digitize the
accepted signals. The FEBs also perform fast analog summing of cell
signals in predefined projective ``towers'' for the L1 trigger. 

The digitized signals are transmitted via
optical fibers to the Readout Drivers (RODs)~\cite{BACKEND_CONS} located in the counting
room 70 m away. The cell energy is reconstructed online in the ROD
modules up to a nominal maximum L1 rate of 75~kHz. The
cell and trigger tower energy reconstruction is described
in detail in Section~\ref{sec:Erec}. 

The response of the 182,468 readout cells is regularly monitored using 122
calibration boards~\cite{CALIB} located in the
front-end crates. These boards inject calibrated current pulses through
high-precision resistors to simulate energy
deposits in the calorimeters. At the end of September 2009, 
1.3\% of cells have problems. The majority of them, i.e. 1.2\% of the
total number of cells, are not read-out because they are connected to
17 non-functioning FEBs. On these FEBs, the active part (VCSEL) of the optical
transmitter to the ROD has failed. This failure, 
occurring at a rate of two or three devices per month, is under intensive
investigation and are expected to be fixed during the next LHC shutdown.
The remaining 0.1\% of cells with problems can be split in three sub-types: incurable cells, i.e. 
cells not responding to the input
pulse (0.02\%), or which are permanently (0.03\%) or sporadically (0.07\%) very noisy. 
The first two types are always masked in the event reconstruction (121
cells), while the sporadically very noisy cells, 
not yet well understood, are masked on an event by event basis. 
For cells which do not receive calibration signals (0.3\%) average
calibration constants computed among neighboring cells are used. 
For cells with non-nominal high voltage (6.1\%) a software correction
factor is applied. Both have very limited impact on the energy
reconstruction.

In total, 180,128 cells, representing 98.7\% of the total number of
cells in the LAr calorimeter, are used for event reconstruction at the end
of September 2009. The number of inactive cells (1.3\%) is dominated
by the cells lost due to faulty optical drivers (1.2\%): apart from
these, the number of inactive cells has been stable in time.

\subsection{Data taking conditions}
\label{sec:Data}

The results presented here focus on the period starting in September 2008 
when all the ATLAS sub-detectors 
were completed and integrated into the data acquisition. 
Apart from regular electronics calibration runs, two interesting types
of data are used to commission the LAr calorimeter: the beam splash
events and the cosmic muons.
The first type corresponds to LHC events of September 10$^{\rm th}$ 
2008 when the first LHC beam hit the collimators located 200~m
upstream of the ATLAS interaction point. 
A cascade of pions and muons parallel to the beam axis fired the 
beam related trigger, illuminated the whole ATLAS detector 
and deposited several PeV per event in the
LAr calorimeter. The second type corresponds to long cosmic muon
runs acquired on September-October 2008 and on June-July 2009
where more than 300 million events were recorded, corresponding to
more than 500 TB of data.

For the LAr commissioning, L1 calorimeter triggers are used to record
radiative energy losses from cosmic muons while the first level
muon spectrometer and second level inner detector triggers are used 
to study pseudo-projective minimum ionizing muons. In most of the runs analyzed, the 
toroidal and solenoidal magnetic fields were at the nominal value.

%% file: NoisePedestal_epj.tex
\section{Electronic performance and quality of cell energy reconstruction}
\label{sec:Erec}

The robustness of the LAr calorimeter energy reconstruction has
been studied in detail using calibration and randomly
triggered events, cosmic muons and beam splash events.
Section~\ref{sec:ErecDef} briefly describes the energy reconstruction
method in the trigger towers and in the cells, as well as a validation
study of the trigger. The time stability of the electronics is
discussed in Section~\ref{sec:noise_ped}. The status of the electronics
timing for the first LHC collisions is presented in
Section~\ref{sec:timing}, and the quality of the LAr calorimeter energy
reconstruction is assessed in Section~\ref{sec:ErecQuality}.  

\subsection{Energy reconstruction in the LAr calorimeter}
\label{sec:ErecDef}

When charged particles cross the LAr gap between electrodes and
absorbers, they ionize the liquid argon. Under the influence of the electric
field, the ionization electrons drift towards the electrode 
inducing a current. The initial current
is proportional to the energy deposited in the liquid argon. The calorimeter
signals are then used to compute the energy per trigger tower or per cell
as discussed in this section.

\subsubsection{Energy reconstruction at the first level calorimeter trigger}
\label{sec:EcellTrig}

The timing requirements for the 
L1 trigger latency can only be met with fast analogue summing in coarse granularity. 
In the EM part, the pre-summation of analog signals
per layer on the FEBs serves as input to tower builder boards where the final 
trigger tower signal sum and shaping is performed. In the HEC and
FCal, the summation is performed on the FEBs 
and transmitted to tower driver boards where only shaping is done.
The tower sizes are $\Delta \eta \times \Delta \phi=0.1\times0.1$ for $|\eta|<2.5$ and 
go up to $\Delta \eta \times \Delta \phi=0.4\times0.4$ for $3.1<|\eta|<4.9$.
The analog trigger sum signals are sent to receiver modules in the service cavern. 
The main function of these modules is 
to compensate for the differences in energy calibration and signal
attenuation over the long cables using programmable amplifier gains ($g_{\rm R}$). The outputs
are sent to L1 trigger pre-processor 
boards which perform the sampling at 40 MHz and the digitization of
five samples. 
At this stage, both the transverse energy and bunch crossing are
determined using a finite impulse response filter, in order to
maximize the signal-to-noise ratio and bunch crossing identification
efficiency. During ATLAS operation, the output $g_{\rm R}A^{\rm L1}$ of the filter, which uses optimal 
filtering, is passed to a look-up table where pedestal ($P$ in ADC counts) subtraction, noise suppression
and the conversion from ADC counts to transverse energy in GeV ($F_{\rm ADC \rightarrow GeV}^{\rm L1}$)
is performed in order to extract the final transverse energy value
($E_{\rm T}^{\rm L1}$) for each trigger tower:
\begin{equation}
E_{\rm T}^{\rm L1} =F_{\rm ADC \rightarrow GeV}^{\rm L1} (g_{\rm R}A^{\rm L1}-g_{\rm R}P).
\label{eq:AmaxTrig}
\end{equation}
Arrays (in $\eta-\phi$) of these $E_{\rm T}^{\rm L1}$ energies, merged with similar information
coming from the Tile calorimeter, are subsequently used to 
trigger on electrons, photons, jets, $\tau$s and events with large missing
transverse energy. 

\subsubsection{Energy reconstruction at cell level}
\label{sec:Ecell}

At the cell level, the treatment of the analog signal is also performed in the
front-end electronics. After shaping, the signal is sampled
at $40$ MHz and digitized if the event was selected by the L1
trigger. The reconstruction of the cell energy, 
performed in the ROD, is based on an optimal filtering algorithm
applied to the samples $s_j$~\cite{OF_REF}. The 
amplitude $A$, in ADC counts, is computed as:
\begin{equation}
A = \sum_{j=1}^{\rm N_{\rm samples}} a_j (s_j-p) \ ,
\label{eq:Amax}
\end{equation}
where $p$ is the ADC pedestal (Section~\ref{sec:pedestal}).
The Optimal Filtering Coefficients (OFCs) $a_j$ are computed per cell
from the predicted ionization pulse shape and the 
measured noise autocorrelation to minimize the
noise and pile-up contributions to $A$.
For cells with sufficient signal, the difference ($\Delta t$ in ns)
between the digitization time and the chosen phase is obtained from:
\begin{equation}
  \Delta t = \frac{1}{A} \sum_{j=1}^{\rm N_{\rm samples}} b_j (s_j-p),
  \label{eq:DeltaT}
\end{equation}
where $b_j$ are time-OFCs. For a perfectly timed detector and in-time
particles $|\Delta t|$ must be close to zero, while larger values
indicate the need for better timing or the presence of out-of-time
particles in the event.

The default number of samples 
used for $A$ and $\Delta t$ computation is ${\rm N_{\rm samples}}=5$, but for some 
specific analyses more samples, up to a maximum of 32, are recorded. 
Finally, including the relevant electronic
calibration constants, the deposited energy (in MeV) is extracted
with: 
\begin{equation}
E_{\rm cell} = F_{\rm \mu A \rightarrow MeV} \times F_{\rm DAC\rightarrow \mu A} \times \frac{1}{\frac{\rm Mphys}{\rm Mcali}} \times G \times A,
\label{eq:Erec}
\end{equation}
where the various constants are linked to the calibration system: the
cell gain $G$ (to cover energies ranging from a maximum of 3 TeV down to noise
level, three linear gains are used: low, medium and high with ratios $\sim
1/10/100$) is computed by injecting a known calibration signal and
reconstructing the corresponding cell response; the factor $1/\frac{\rm Mphys}{\rm Mcali}$
quantifies the ratio of response to a calibration pulse and an 
ionization pulse corresponding to the same input current; 
the factor $F_{\rm DAC\rightarrow \mu A}$ converts 
digital-to-analog converter (DAC) counts set on the calibration board to $\mu A$; 
finally, the factor $F_{\rm \mu A\rightarrow MeV}$ is estimated from
simulations and beam test results, and includes high
voltage corrections for non-nominal settings
(see Sec~\ref{sec:CaloHV}). Note that the crosstalk bias in the finely segmented first layer of
the electromagnetic calorimeter is corrected for in the gain
$G$~\cite{ATLAS_TB_EM3}.

\subsubsection{Check of the first level tower trigger energy computation}
\label{sec:L1Calo}

The trigger decision is of utmost importance for ATLAS
during LHC collisions since the data-taking rate is at maximum
200 Hz because of bandwidth limitations, i.e. a factor $2\times10^5$ smaller than
the 40 MHz LHC clock. It is therefore important to check that no
systematic bias is introduced in the computation of the L1
trigger energy and that the trigger energy resolution is not too degraded
with respect to the offline reconstruction. In the following, this
check is performed with the most granular part of the LAr calorimeter,
the barrel part of the EM calorimeter, where 60 cell signals are
summed per trigger tower. 

Since cosmic muon events occur asynchronously with respect to the LHC clock,
and the electronics for both the trigger and the standard readout is loaded with one
set of filtering coefficients (corresponding to beam crossing), the reconstructed
energy is biased by up to 10\%, depending on the phase. 
For the study presented here, $A^{\rm L1}$ is  
recomputed offline by fitting a second-order polynomial to the three highest samples 
transmitted through the processors. The
most critical part in the trigger energy computation is then to calibrate the individual 
receiver gains $g_{\rm R}$. For that purpose, a common linearly increasing
calibration pulse is sent to both the L1 trigger and the normal cell circuits: the inverse
receiver gain $1/g_{\rm R}$ is obtained by fitting the
correlation between the L1 calorimeter transverse energy
($E_{\rm T}^{\rm L1}$) and the sum of cell transverse energies in the same trigger
tower, later called {\it offline} trigger tower ($E_{\rm T}^{\rm LAr}$). In cosmic muon runs, receiver gains are set to 1.0
and are recomputed offline with dedicated calibration runs. 
As a cross check, the gain was also extracted
using LHC beam splash event data which covers the full detector. In both cases, 
the L1 transverse energy is computed as in Eq.~\ref{eq:AmaxTrig}. 

\begin{figure}[hbtp]
\begin{center}
\includegraphics[width=0.45\textwidth]{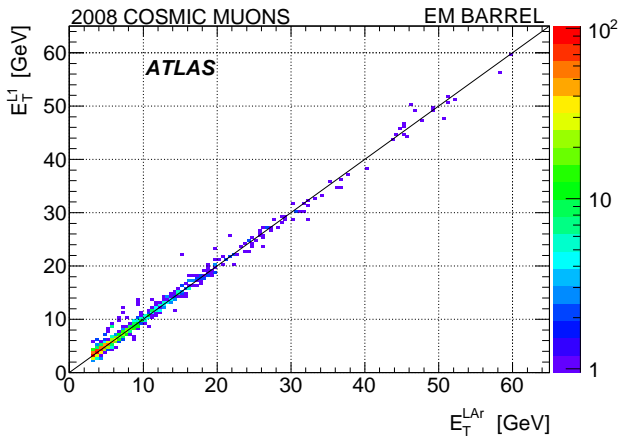}
\end{center}
\caption{\it L1 transverse energy ($E_{\rm T}^{\rm L1}$)
  computed with the receiver gains extracted from calibration runs
  versus the sum of cell transverse energies in the same trigger tower
  ($E_{\rm T}^{\rm LAr}$). }
\label{fig:L1CaloCal1}
\end{figure}

In the EM calorimeter, radiating cosmic muons may produce a local energy deposit
of a few GeV, and fire the EM calorimeter trigger condition EM3 that
requires a transverse energy greater than 3 GeV in a sum of four adjacent EM trigger
towers. To mimic an electron coming from the interaction point, only those events that
contain a track reconstructed with strict projectivity cuts are
considered. Here, the L1 calorimeter 
transverse energy is computed using the gains determined with calibration
runs. Figure~\ref{fig:L1CaloCal1} shows the correlation between
$E_{\rm T}^{\rm L1}$ and $E_{\rm T}^{\rm LAr}$. Computing the ratio
of $E_{\rm T}^{\rm L1}$ and $E_{\rm T}^{\rm LAr}$ gives a Gaussian distribution with a mean of 1.015$\pm$0.002,
showing the very good correspondence between these two quantities, especially at low energy. This also shows that 
the trigger energy is well calibrated and almost unbiased with respect to the LAr readout. 

\begin{figure}[hbtp]
\begin{center}
\includegraphics[width=0.45\textwidth]{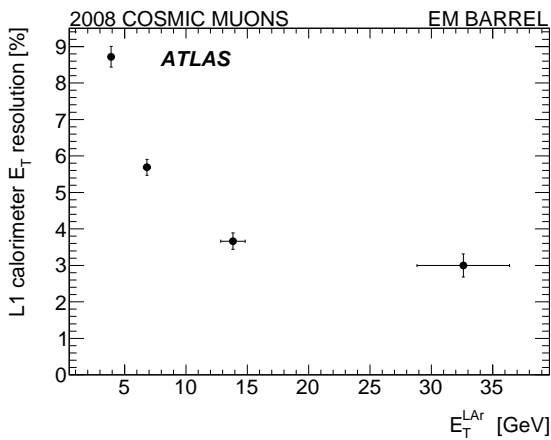}
\end{center}
\caption{\it Relative difference of $E_{\rm T}^{\rm L1}$ and $E_{\rm T}^{\rm LAr}$ (L1 Calorimeter $E_{\rm T}$ 
  resolution) as a function of 
  $E_{\rm T}^{\rm LAr}$. Strict projectivity cuts for the track pointing to
  the EM shower are applied. Horizontal error bars reflect the RMS of $E_{\rm T}^{\rm LAr}$ in each bin. }
\label{fig:L1CaloCal2}
\end{figure}

Figure~\ref{fig:L1CaloCal2} shows the corresponding
resolution computed as the relative difference of $E_{\rm T}^{\rm L1}$ and
$E_{\rm T}^{\rm LAr}$. At low energy, the difference is dominated by
electronic noise since the two readout paths have only part of
their electronics in common. The ATLAS specification of 5\% of L1 transverse 
energy resolution is reached for energies greater than 10 GeV. The L1 
transverse energy resolution reaches around 3\% at high energy. 

\begin{figure}[hbtp]
\begin{center}
\includegraphics[width=0.45\textwidth]{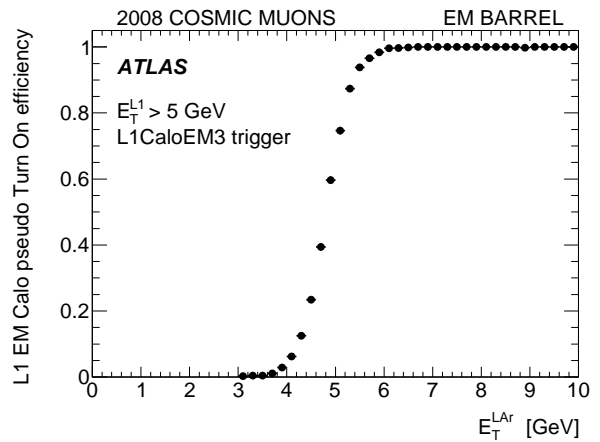}
\end{center}
\caption{\it Turn-on curve efficiency for
  $E_{\rm T}^{\rm L1}>5$GeV requirement obtained with events 
  triggered by the EM3 L1 Calorimeter trigger.} 
\label{fig:L1CaloBS}
\end{figure}

As a crosscheck, a similar study was performed with gains computed
from the beam splash events, without the projectivity cut. A slight
degradation of the resolution is observed at high energy, but not at
low energy where the noise dominates. Taking advantage of the higher
statistics, it is possible to compute the 5~GeV ``turn-on curve'', i.e. 
the relative efficiency for an offline trigger tower to meet
the requirement $E_{\rm T}^{\rm L1}\geq 5$ GeV as a function of
$E_{\rm T}^{\rm LAr}$. This is not the absolute efficiency as the
calorimeter trigger condition EM3 is used to trigger the events. The efficiency is shown in
Figure~\ref{fig:L1CaloBS}, where a sharp variation around a $E_{\rm
  T}^{\rm L1}=5$ GeV energy threshold is observed.  

These results give confidence that EM showers (electrons and photons)
will be triggered efficiently in LHC
events. After this study, the gains $g_{\rm R}$ 
were extracted 
from dedicated calibration runs and loaded into the receivers to be used for the 
first LHC collisions.

\subsection{Electronic stability}
\label{sec:noise_ped}

Hundreds of millions of randomly triggered and calibration events 
can be used for a study of the stability of the properties of each readout channel,
such as the pedestal, noise and gain. The first two quantities are computed for 
each cell as the mean (pedestal) of the signal samples $s_j$ in 
ADC counts, and the width (noise) of the energy distribution. The gain is extracted 
by fitting the output pulse amplitudes against calibration pulses with
increasing amplitudes.

\subsubsection{Pedestal}
\label{sec:pedestal}

The stability of the pedestals is monitored by measuring variations 
with respect to a reference pedestal value for each cell. For each FEB, 
an average over the 128 channels is computed. 

\begin{figure}[hbtp]
\begin{center}
\includegraphics[width=0.45\textwidth]{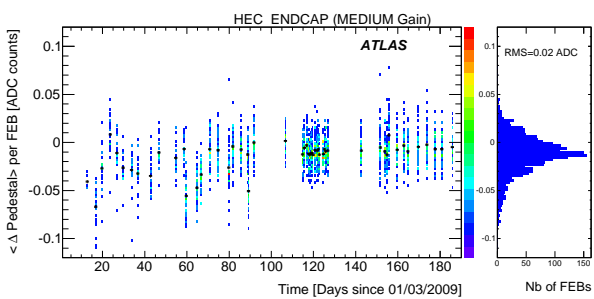}
\end{center}
\caption{\it Average FEB pedestal variations in ADC counts, in medium gain, for the HEC during 6
  months of data taking in 2009. The crosses indicate the mean value for 
each time slice.} 
\label{fig:pedestal}
\end{figure}

As an example, Figure~\ref{fig:pedestal} illustrates the results for
the 48 HEC FEBs over a period of six months in 2009. A slight drift
of the pedestal with time, uncorrelated with the FEB temperature
and/or magnetic field configurations, is observed. Overall, the FEB
pedestal variations follow a Gaussian distribution with a standard
deviation of 0.02 ADC counts, i.e. below 2 MeV. The same checks have
been performed on all other FEBs, and give typical variations of
around 1 (0.1) MeV and 10 (1) MeV in the EM and FCal calorimeters 
respectively, in medium (high) gain. These variations are much lower for
the EM and HEC or at the same level for the FCal than the
numerical precision of the energy computation, which is 8 (1)
MeV in medium (high) gain. 

During the LHC running, it is foreseen to acquire pedestal and calibration 
runs between fills, thus it will be possible to correct for any small time dependence 
such as observed in Figure~\ref{fig:pedestal}. In the same spirit, random triggers
collected during physics runs can be used to track any pedestal variations
during an LHC fill. 

\subsubsection{Noise}
\label{sec:noise}

\begin{figure}[hbtp]
\begin{center}
\includegraphics[width=0.45\textwidth]{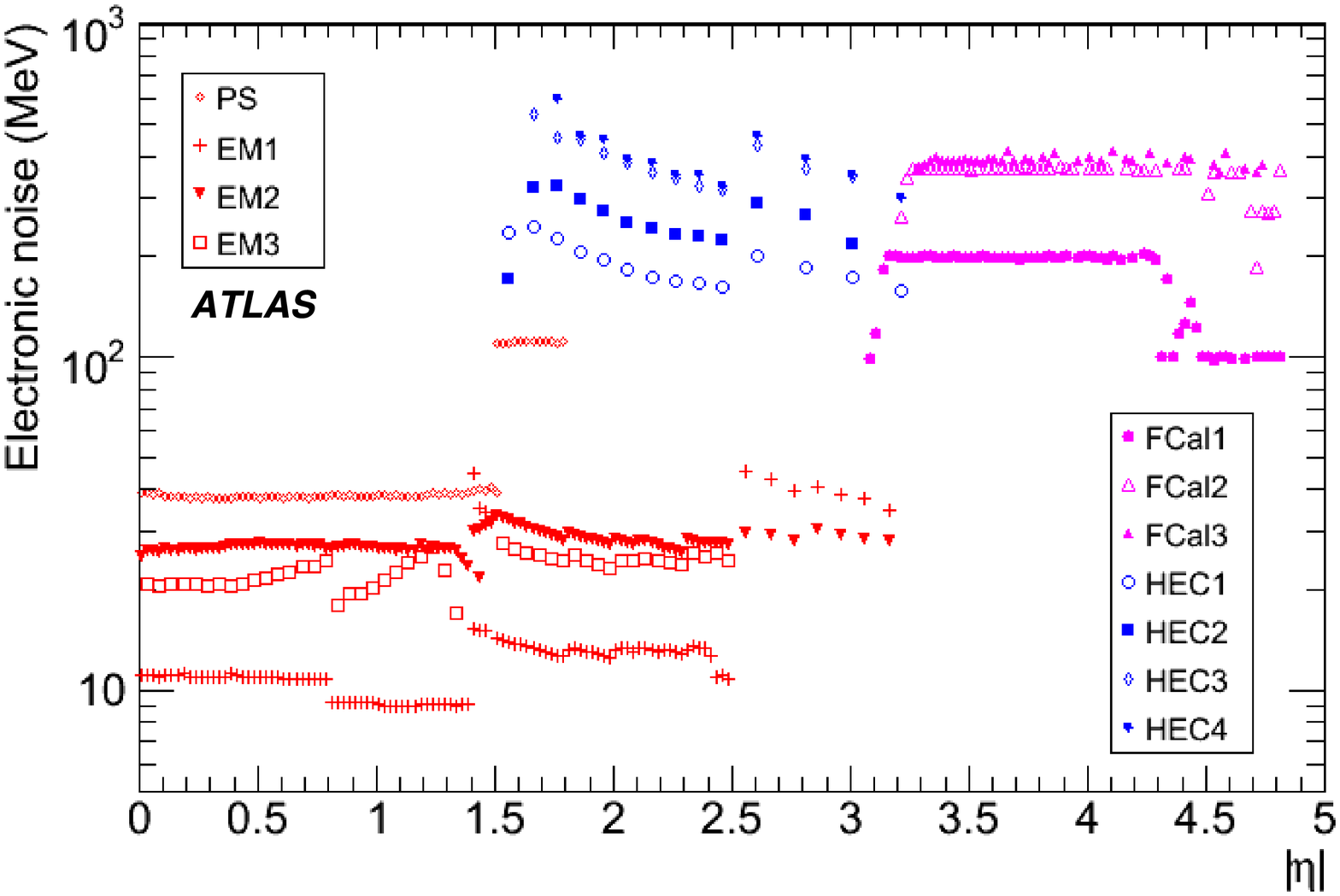}
\end{center}
\caption{\it Electronic noise ($\sigma_{\rm noise}$) in randomly
  triggered events at the EM scale in
  individual cells for each layer of the
  calorimeter as a function of $|\eta|$. Results are averaged over $\phi$.} 
\label{fig:noise}
\end{figure}

Figure~\ref{fig:noise} shows the noise measured in randomly triggered events at 
the cell level as a function of $\eta$ for all layers of the LAr
calorimeters. In all layers, a good agreement with the expected noise~\cite{ATLAS_PHYS_TDR} is 
observed. Noise values are symmetric with respect to $\eta=0$ and uniformly in $\phi$ within few 
percents. In the EM calorimeters, the noise ranges from 10 to 50~MeV,
while it is typically a factor of 10 greater in the hadronic endcap
and forward calorimeters where the granularity is 20
times coarser and the sampling fractions are lower. It should be noted
that these results are obtained 
using five samples in Eq.~(\ref{eq:Amax}) and~(\ref{eq:Erec}), i.e. the noise is reduced
by a factor varying from 1.5 to 1.8, depending on $\eta$, with respect
to the single-sample noise value. 

The coherent noise over the many cells used to measure electron and
photon energies in the EM calorimeters should be kept below 5\%~\cite{LARG_TDR} 
of the incoherent noise (i.e. the quadratic sum of all channel noise). 
For the second layer of the EM calorimeter, the contribution from the
coherent noise has been estimated to $2\%$, by studying simultaneous
increase of noise in a group of channels.  

Systematic studies of noise stability have been pursued: all noise variations
are typically within $\pm$ 1 keV, 0.1 MeV and 1 MeV for EM, HEC and FCal, respectively. No
correlations with the FEB temperature and/or changes of magnetic field
conditions have been observed.  

\subsubsection{Gain}
\label{sec:gain}

The calibration pulse is an exponential signal (controlled by two
parameters, $f_{\rm step}$ and $\tau_{\rm cali}$) which emulates the triangular
ionization signal. It is injected on the detector as close as possible to the
electrodes, except for the FCal where it is
applied at the base-plane of the front-end crates~\cite{FCAL_CONS}. Thus, 
the analog cell response is
treated by the FEBs in the same way as an ionization signal, but it
is typically averaged over 100 triggers in the RODs and transmitted
offline where the average signal peak height is computed. The cell gain
is extracted as the inverse ratio of the response signal in ADC counts to the
injected calibration signal in DAC counts.

\begin{figure}[hbtp]
\begin{center}
\includegraphics[width=0.45\textwidth]{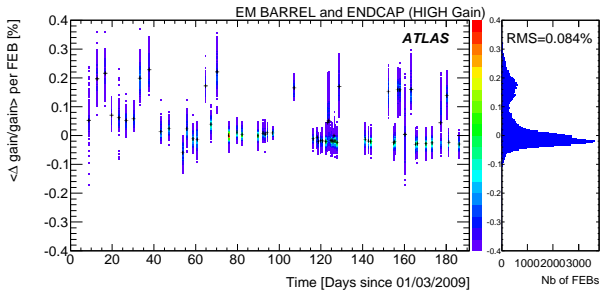}
\end{center}
\caption{\it Average FEB (high) gain variations during 6 months
  of 2009 data taking, in the EM part of the calorimeter. The crosses indicate the mean value for
each time slice.} 
\label{fig:calib}
\end{figure}

The stability of the cell gain is monitored by looking at the relative
gain difference averaged over 128 FEB channels. This is illustrated in
Figure~\ref{fig:calib} for the 1448 FEBs of the EM
calorimeter, in high gain. All variations are within $\pm 0.3\%$ and
similar results are obtained for medium and low gains. 
An effect of $0.2\%$ on the gains has recently been identified as coming 
from a particular setting of the FEBs. The two populations are most 
probably coming from this effect. Regular update of calibration database 
take account of the variations.
Similar results are obtained for the HEC, and variations within $\pm 0.1\%$ 
are measured for the FCal.

In conclusion, results presented for the pedestals, noise, and 
gains illustrate the stability of the LAr electronics over
several months of data taking. Values are stored in the ATLAS calibration
database and are used for online and offline reconstruction. 

\subsubsection{Global check with \etmiss~variable}
\label{sec:etmiss}

Another way to investigate the level of understanding of pedestals and
noise in the LAr calorimeter is to compute global quantities in randomly triggered events
with the calorimeter, such as the vector sum of transverse
cell energies. The calorimetric missing transverse energy \etmiss~is
defined as:   
\begin{equation}
\begin{array}{l}
E_{\rm x}^{\rm miss}\,=\,-\sum_{i=1}^{N_{\rm cell}} E_i \sin\theta_i \cos\phi_i, \\
E_{\rm y}^{\rm miss}\,=\,-\sum_{i=1}^{N_{\rm cell}} E_i \sin\theta_i \sin\phi_i,\\
E_{\rm T}^{\rm miss}\,=\,\sqrt{(E_{\rm x}^{\rm miss})^2+(E_{\rm y}^{\rm miss})^2}, \\
\end{array}
\label{eq:etmiss}
\end{equation}
where $E_i$ is the cell energy, $\theta_i$ its polar angle and $\phi_i$ its azimuthal angle.
Because of the high granularity of the LAr calorimeter, it is
crucial to suppress noise contributions to \etmiss, i.e. limit the number of cells, 
$N_{\rm cell}$, used in the sum. In ATLAS, this is done with two methods: 
$i)$ a cell-based method in which only cells above
a noise threshold of two standard deviations ($|E_i|>2\sigma_{\rm noise}$) are kept;
$ii)$ a cluster-based method which uses only cells belonging to three-dimensional topological
clusters~\cite{clusters}. These clusters are built around $|E_i|>4\sigma_{\rm noise}$ seeds by
iteratively gathering neighboring cells with $|E_i|>2\sigma_{\rm noise}$ and, in a final step, 
adding all direct neighbors of these accumulated secondary cells (Topocluster 4/2/0). In 
randomly triggered events, about 8500 and 500 LAr cells, respectively, are selected with these two
noise-suppression methods. 

The distributions of $E_{\rm x}^{\rm miss}$ and $E_{\rm y}^{\rm miss}$ should be Gaussian 
and centered on zero in randomly triggered events. The measurements are compared 
with a Gaussian noise model, where no pedestal shift or coherent noise is present,
obtained by randomizing the cell energy according to a Gaussian model for the cell
noise. For this \etmiss~computation, cells with very high noise (see
Section~\ref{sec:CaloReadOut}) are removed from the computation. 

\begin{figure}[hbtp]
\begin{center}
\includegraphics[width=0.45\textwidth]{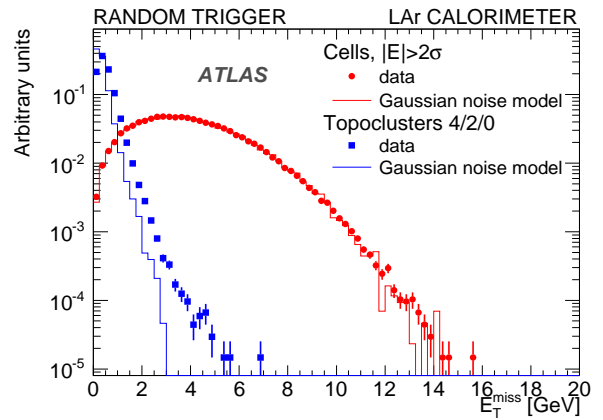}
\end{center}
\caption{\it \etmiss~distribution with LAr calorimeter cells 
  for 135,000 randomly triggered events in June 2009. The dots (squares) show the 
  cell-based (cluster-based) methods in the data, and the
  histograms show the equivalent distributions for the Gaussian noise model (see text).}
\label{fig:etmiss1}
\end{figure}

Figure~\ref{fig:etmiss1} shows the \etmiss~distributions for a randomly 
triggered data sample acquired in 15 hours. The two noise suppression methods 
are compared to the corresponding Gaussian noise model. For the
cell-based method, a good agreement is observed between the data and the simple model. 
Because of the lower number of cells kept in the cluster-based method, 
a smaller noise contribution to $E_{\rm T}^{\rm  miss}$ is observed. 
The agreement between the data and the model is not as good as for
the cell-based method, reflecting the higher
sensitivity of the cluster-based method to the noise description. In both cases, no  
\etmiss~tails are present, reflecting the absence of large systematic pedestal shifts or
abnormal noise. 

Using \etmiss~it was possible to spot, in
2008, a high coherent noise due to the defective grounding of a
barrel presampler HV cable and sporadic noise in a 
few preamplifiers. These two problems were repaired prior to 
the 2009 runs. The time stability of
\etmiss~is regularly monitored using randomly triggered events by
observing the mean and width of the $E_{\rm x}^{\rm miss}$ and 
$E_{\rm y}^{\rm miss}$ distributions. 
With the cluster-based method, the variation
of all quantities was measured to be $\pm0.1$ GeV
over 1.5 months. This variation is small compared to the expected
\etmiss~resolution ($\simeq 5$ GeV for $W\to e \nu$ events) and can be
controlled further by more frequent updates of the calibration constants. 

\begin{figure}[hbtp]
\begin{center}
\includegraphics[width=0.45\textwidth]{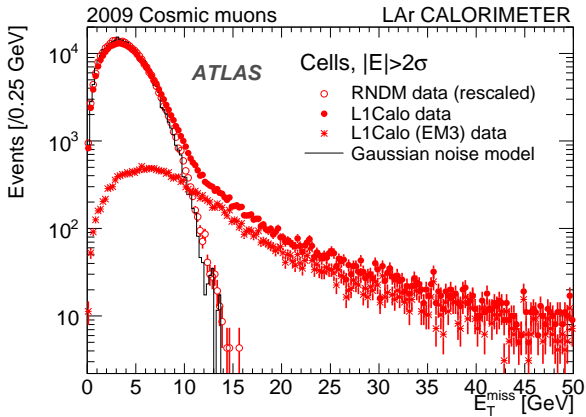}
\end{center}
\caption{\it \etmiss~distribution with LAr calorimeter cells for 300,000 L1 calorimeter (L1Calo) triggers
  reconstructed with the cell-based method. Results for EM3 trigger 
  conditions (Section~\ref{sec:L1Calo}) from the same run are superimposed 
on the same plot and the results from randomly 
  triggered events are again overlaid (open symbols and histogram).}  
\label{fig:etmiss2}
\end{figure}

A similar analysis was performed with L1 calorimeter triggered events,
corresponding to radiative energy losses from cosmic muons,
from the same run as used above. 
The L1 calorimeter trigger (L1calo) triggers events when either  the
sum of adjacent trigger tower transverse energies is above 3 GeV in
the EM calorimeter (EM3) or 5 GeV when summing EM and hadronic
towers~\cite{L1Calo}. 
The results are
illustrated in Figure~\ref{fig:etmiss2} for the cell-based noise
suppression method. Most of these events are triggered by energy losses in the
Tile calorimeter that do not spill in the LAr calorimeter, which therefore
mainly records noise, leading to a \etmiss~distribution similar to
the one obtained with random triggers. However, in few cases, events
are triggered by the LAr calorimeter such as the EM3 trigger. The
peak at 3~GeV is then shifted upwards to 6~GeV and the proportion of
events with \etmiss~above 15 GeV is greatly enhanced. 

%% file: Timing_epj.tex
\subsection{LAr calorimeter timing}
\label{sec:timing}

The energy reconstruction in each cell relies on the fact that in the
standard (five samples) physics data acquisition mode, the third sample is
located close to the signal maximum: this implies an alignment of the
timing of all calorimeter cells to within a few ns. 

Several parameters determine each cell timing: the first contribution
comes from FEB internal delays which induce a cell timing
variation of $\pm 2$ ns within each FEB. This is
accounted for when computing the optimal filtering coefficients. The
second contribution concerns FEB to FEB variations due to different
cable lengths to reach a given FEB: this relative FEB
timing can vary by up to $\pm 10$ ns and can be corrected for by
setting an adjustable delay on each FEB. 

The study presented here aims at predicting (using calibration data
and additional hardware inputs) and measuring (using cosmic muons and
beam splash data) this relative FEB timing in order to derive timing
alignment delays for each FEB.  

\subsubsection{Timing prediction}
\label{sec:timing_pred}

The time of the signal maximum is different in a
calibration run ($t_{\rm calib}$) and in a physics run ($t_{\rm
  phys}$). The main contribution to this time is the delay $T_{\rm 0}$
before the pulse starts to rise (the difference between the
calibration and physics pulse widths is much smaller than this
$T_{\rm 0}$ delay variation). This delay is driven by cable lengths
which are different in these two configurations and additional delays
in physics runs because of the particle time of flight, and the
Timing, Trigger and Control (TTC) system configurations.    

In a calibration run, a signal is injected from the
calibration board through the calibration cables, and is then read out
through the physics signal cables. The value of the delay $T_{\rm
  0}^{\rm calib}$ with respect to the signal injection can thus be
computed for each FEB using the various  
cable lengths ($L_{\rm calib},~ L_{\rm physics}$) and signal propagation speeds
($v_{\rm calib},~ v_{\rm physics}$): 
\begin{equation}
T_{\rm 0}^{\rm calib} = \frac{L_{\rm calib}}{v_{\rm calib}} + \frac{L_{\rm
  physics}}{ v_{\rm physics}}.
\end{equation}
The above prediction is compared with the measured value in calibration
runs. The measurement corresponds to the time at which the calibration
pulse exceeds three standard deviations above the noise; it
is found to agree with the prediction to within $\pm 2$ ns, ignoring
the variations within each FEB. 

The time of the signal maximum $t_{\rm calib}$ is obtained by fitting
the peak of the pulse of cells in a given FEB with a third order polynomial. As
the cable length is a function of the cell position along the beam axis ($z$,
$\eta$), the cell times are averaged per FEBs in a 
given layer (except for the HEC where layers are mixed inside a FEB)
and a given $\eta$-bin in order to align the FEBs in time.  

The time of the ionization pulse in each cell can then be predicted from
the calibration time using the following formula:
\begin{equation}
\label{eq:physics_pred}
t_{\rm phys} = t_{\rm calib} - \frac{L_{\rm calib}}{v_{\rm calib}} + t_{\rm flight} + \Delta t_{\rm TTC},
\end{equation}
where $t_{\rm calib}$ was defined in the previous paragraph; $t_{\rm flight}$ is the
time of flight of an incident particle from the interaction point to 
the cell, which varies from 5 ns for a presampler cell
at $\eta=0$, to 19 ns for a back cell in the HEC; and $\Delta t_{\rm TTC}$ is a
global correction for the six partitions due to the cabling of the 
TTC system which is needed to align all FEBs at the crate level.  
This predicted ionization pulse time is compared with the
corresponding measurement in the next section. 

\subsubsection{Timing measurement }
\label{sec:timingphysics}

The ionization pulse time has been measured in beam splash and cosmic
muon events. The time is reconstructed using optimal filtering
coefficients. Since the arrival time of the particle is not known, one
does not know in advance to which samples the time OFCs $b_i$ should
be applied (since these OFCs were computed for a particular set of
samples around the pulse maximum). Therefore, an iterative procedure
is used until the obtained $\Delta t$  (see Eq.~\ref{eq:DeltaT}) is
less than $3$ ns.

The time is
then corrected for two effects: first, the time-of-flight difference
between the beam splash or cosmic muon configurations and the collision
configuration, and second, the asynchronicity of the beam splash and
cosmic muon events, where arrival times vary with
respect to the TTC clock.  

\begin{figure*}[hbt]
\begin{center}
\begin{tabular}{cc}
\mbox{\epsfxsize=0.42\hsize\epsfbox{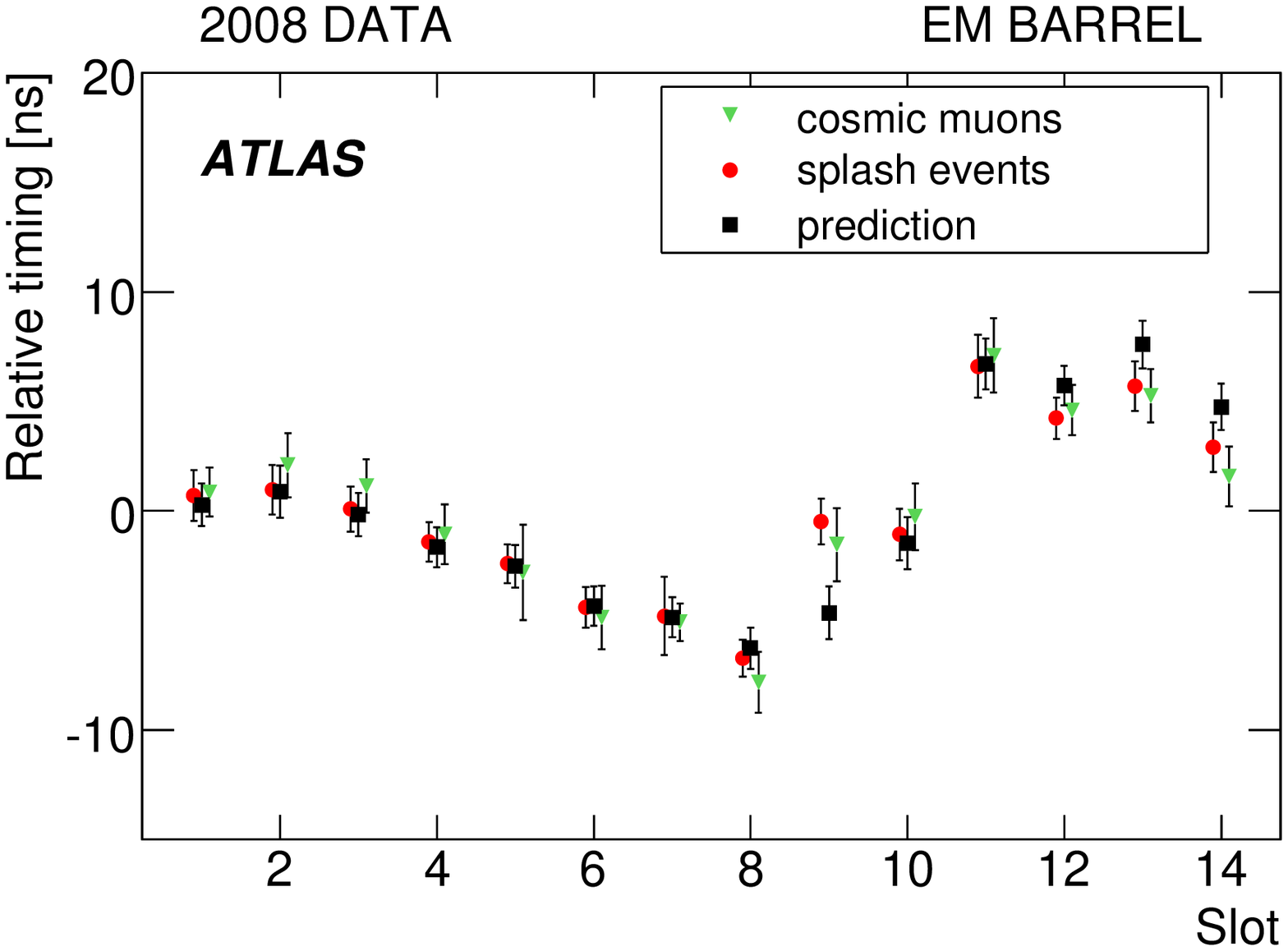}} &
\mbox{\epsfxsize=0.42\hsize\epsfbox{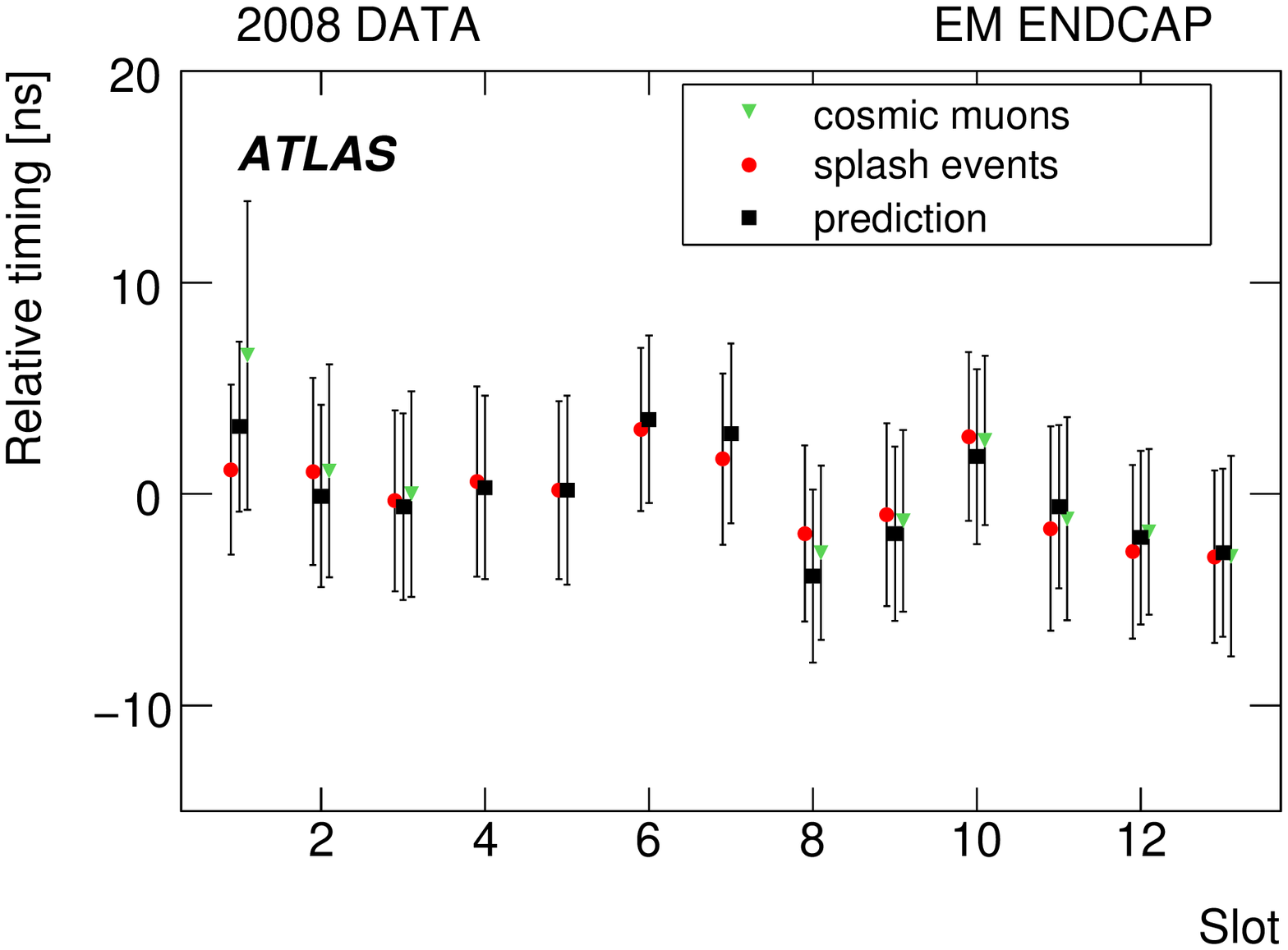}} \\
\mbox{\epsfxsize=0.42\hsize\epsfbox{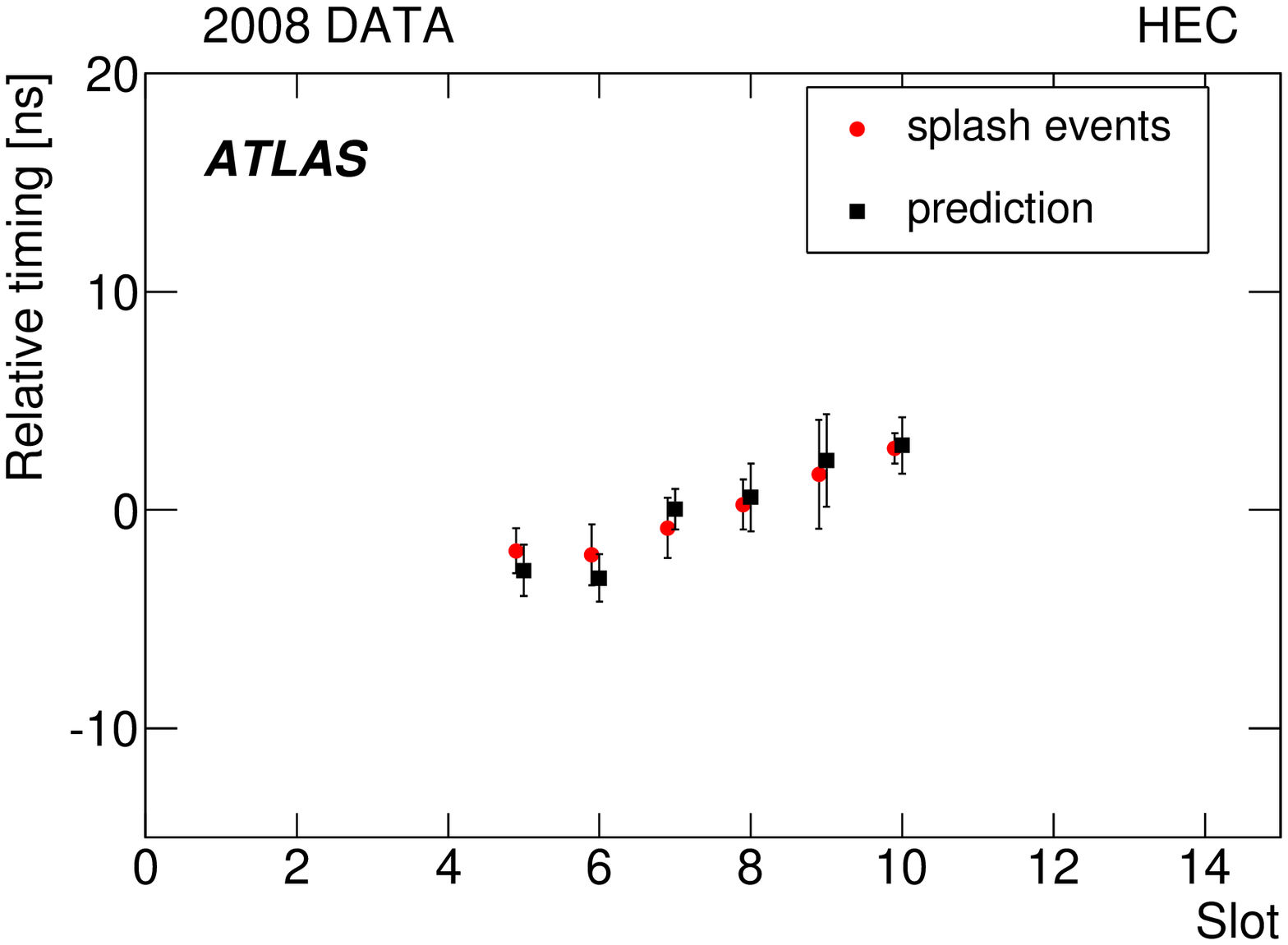}} &
\mbox{\epsfxsize=0.42\hsize\epsfbox{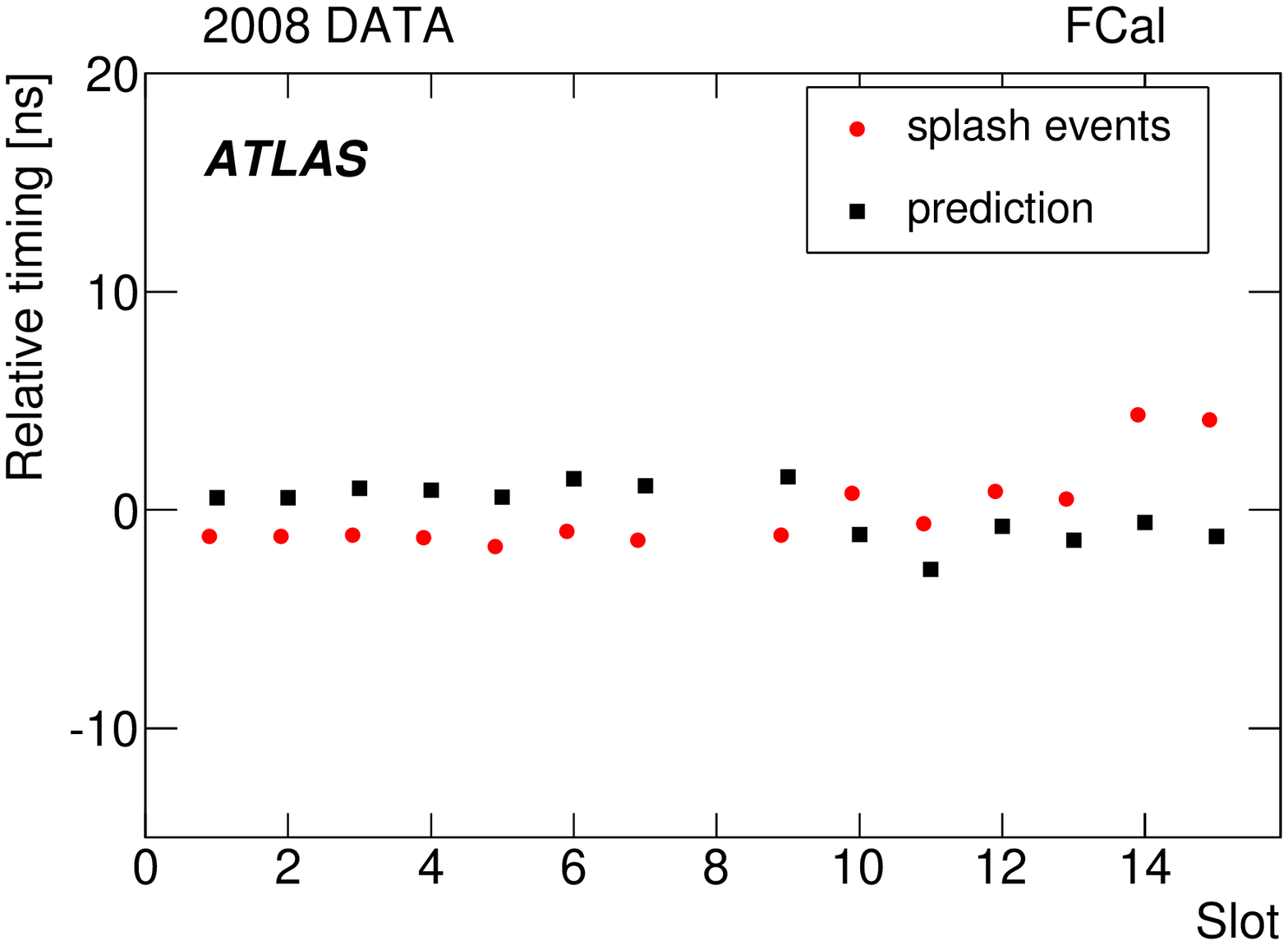}}
\end{tabular}
\caption{\em Relative predicted and measured FEB times in the
  electromagnetic barrel (top left), electromagnetic endcap (top
  right), HEC (bottom left) and FCal (bottom right) calorimeters, for
  the C-side ($\eta<0$). The x-axis (``Slot'') corresponds to a
  group of FEBs in a given layer (or a group of layers in the HEC) and
  $\eta$-range. The error bars show 
  the width of the distributions in each slot.} 
\label{fig:timing}
\end{center}
\end{figure*}

The comparison between the measured and the predicted
(Eq.~\ref{eq:physics_pred}) ionization pulse time is shown in
Figure~\ref{fig:timing} for the C-side ($\eta<0$) of each LAr
sub-detector. 

This comparison is performed for each ``slot'' corresponding to a group of
FEBs in a given layer and $\eta$-range, averaged over all calorimeter
modules over $\phi$. 
As mentionned in the introduction, the relative timing of each group
of FEBs varies by $\pm 10$ ns due to the different corresponding cable
lengthes.  

On the plots, the error bars correspond to 
the RMS of values for all modules in a slot: in the FCal, there
is only one module per slot, so no error bars are shown (also
note that slot 8 is empty in the FCal). In some regions,
the cosmic data statistics was not sufficient to extract the time: the
corresponding bins are thus empty. The agreement
between the prediction and the two measurements is within $\pm 2$ ns
(and at worst $\pm 5$ ns for two slots of the FCal). 

Finally, a set of FEB timing alignment delays is obtained from these
well understood measured relative times. These delays 
will be used at the LHC startup and updated once the phase between the
beam and the machine clock will be measured and shown to be
stable. The desired precision of $\pm 1$ ns should be reached then.

%% file: ErecQuality_epj.tex
\subsection{Signal reconstruction studies and impact on intrinsic global energy resolution constant term}
\label{sec:ErecQuality}

The main ingredient for accurate energy and time reconstruction of signals from 
LHC collisions is the prediction of the ionization signal shape, from
which the optimal filtering coefficients used in
Eq.~(\ref{eq:Amax}) are computed. After recalling the basics of 
the method used to predict the shape
in Section~\ref{sec:PredSigRec}, an estimate of the signal prediction
quality with three samples in the EM calorimeter is presented 
in Section~\ref{sec:QualSigRec}. The full 32 samples 
shape prediction is used to determine the ionization electron drift
time needed for the OFC computation in the EM calorimeter 
(Section~\ref{sec:tdrift}). Finally, from these two studies an 
estimate of the main contributions to the constant term in the global
energy resolution of the EM calorimeter is given in Section~\ref{sec:CT}.  

\subsubsection{Prediction of the ionization pulse shape}
\label{sec:PredSigRec}

The standard ATLAS method for prediction of the ionization pulse
shape in the EM and the HEC relies on the calibration system. A
precisely known calibration signal is sent through the same path as seen by
the ionization pulses thus probing the actual electrical and readout properties of each 
calorimeter cell. In both the EM and the HEC, the calibration pulse properties are 
parameterized using two variables, $f_{\rm step}$ and $\tau_{\rm cali}$, which have been 
measured for all calibration boards~\cite{CALIB} and are routinely extracted 
from calibration signals~\cite{RTM}.

The predicted ionization shapes are calculated
from the calibration pulses by modeling each readout cell as a resonant $RLC$ 
circuit, where $C$ is the cell capacitance, $L$ the inductive path of 
ionization signal, and $R$ the contact resistance
between the cell electrode and the readout line. The effective $LC$ and $RC$ 
have been estimated from a frequency analysis of the output
calibration pulse shape~\cite{RTM}. They were also 
measured with a network analyzer during the long validation
period of the three cryostats~\cite{MEAS,SHAPE0,SHAPE1}. For the HEC,
calibration pulses are transformed into 
ionization signal predictions using a semi-analytical model of the readout 
electronics, with a functional form 
with zeros and poles accounting for the cable and 
pre-amplifier transfer functions~\cite{HECPRED,HECPRED2}.
The prediction of both the EM and HEC ionization pulses requires the 
knowledge of the electron drift time in liquid argon 
($T_{\rm drift}$), which can be inferred from the calorimeter properties or 
directly measured from data (see Section~\ref{sec:tdrift}).
  
\begin{figure*}[htb]
\begin{center}
\begin{tabular}{lr}
\epsfig{file=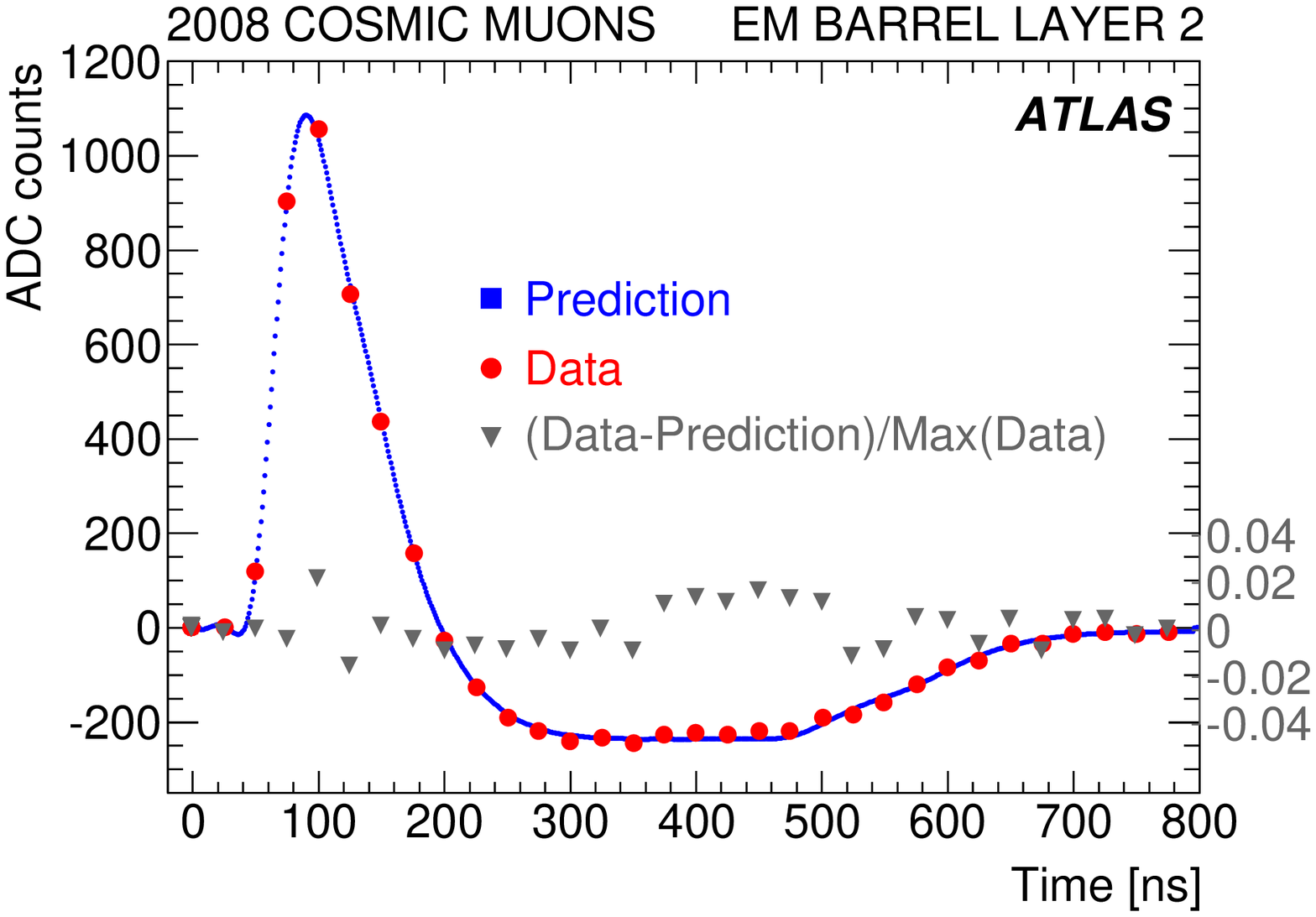,width=.49\linewidth, height=4.9cm} & 
\epsfig{file=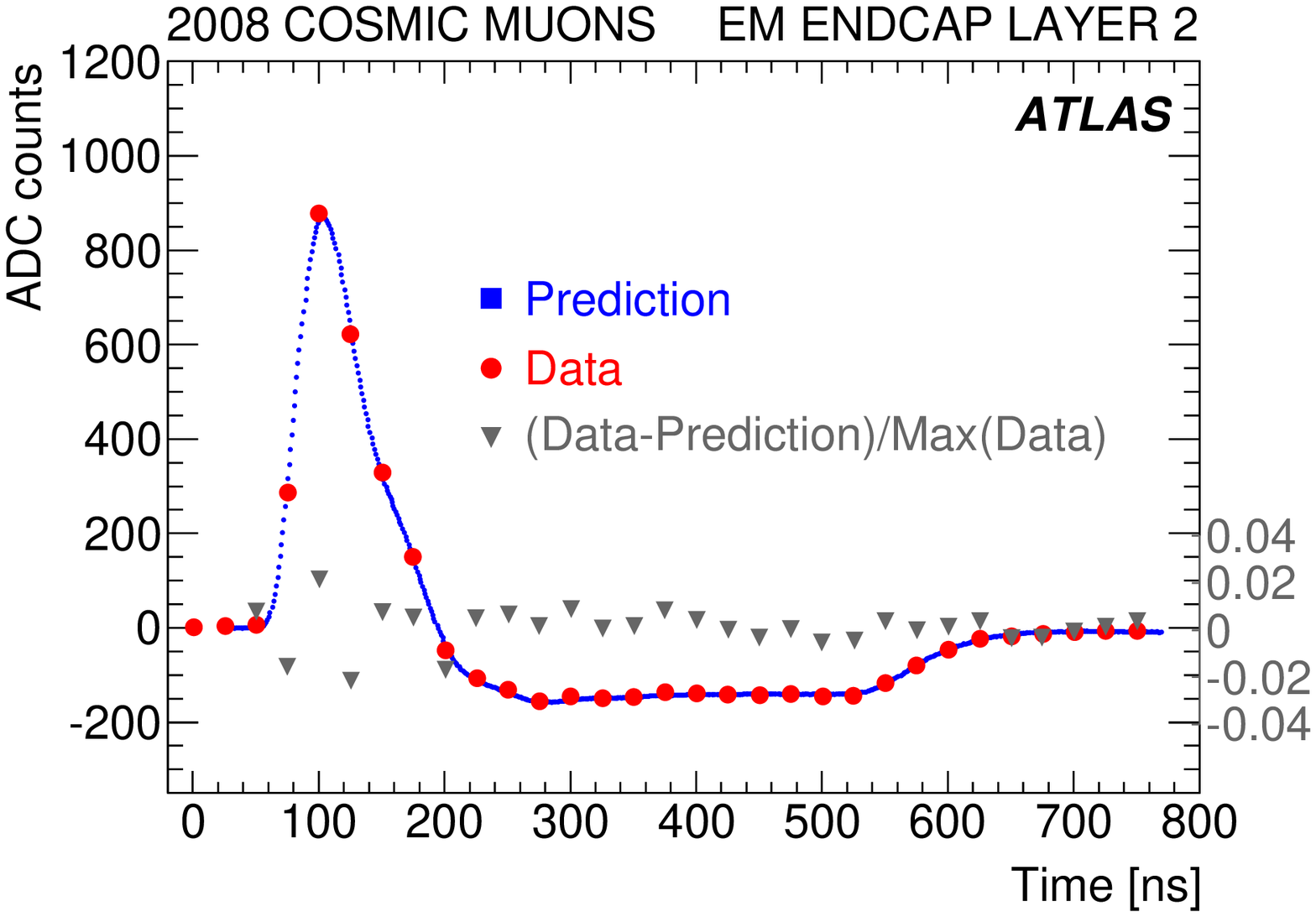,width=.49\linewidth, height=4.9cm} \\
\epsfig{file=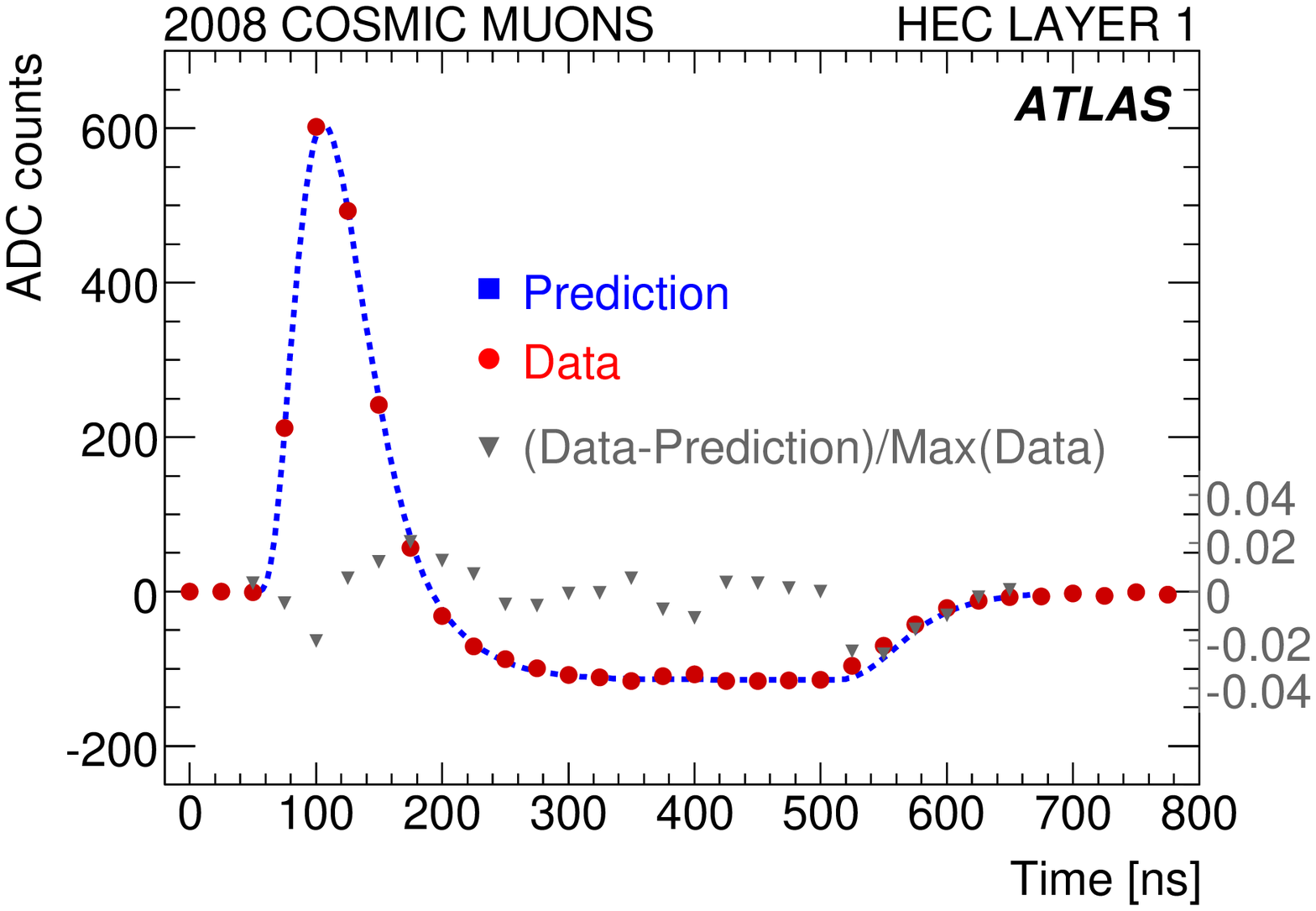,width=.49\linewidth, height=4.9cm} & 
\epsfig{file=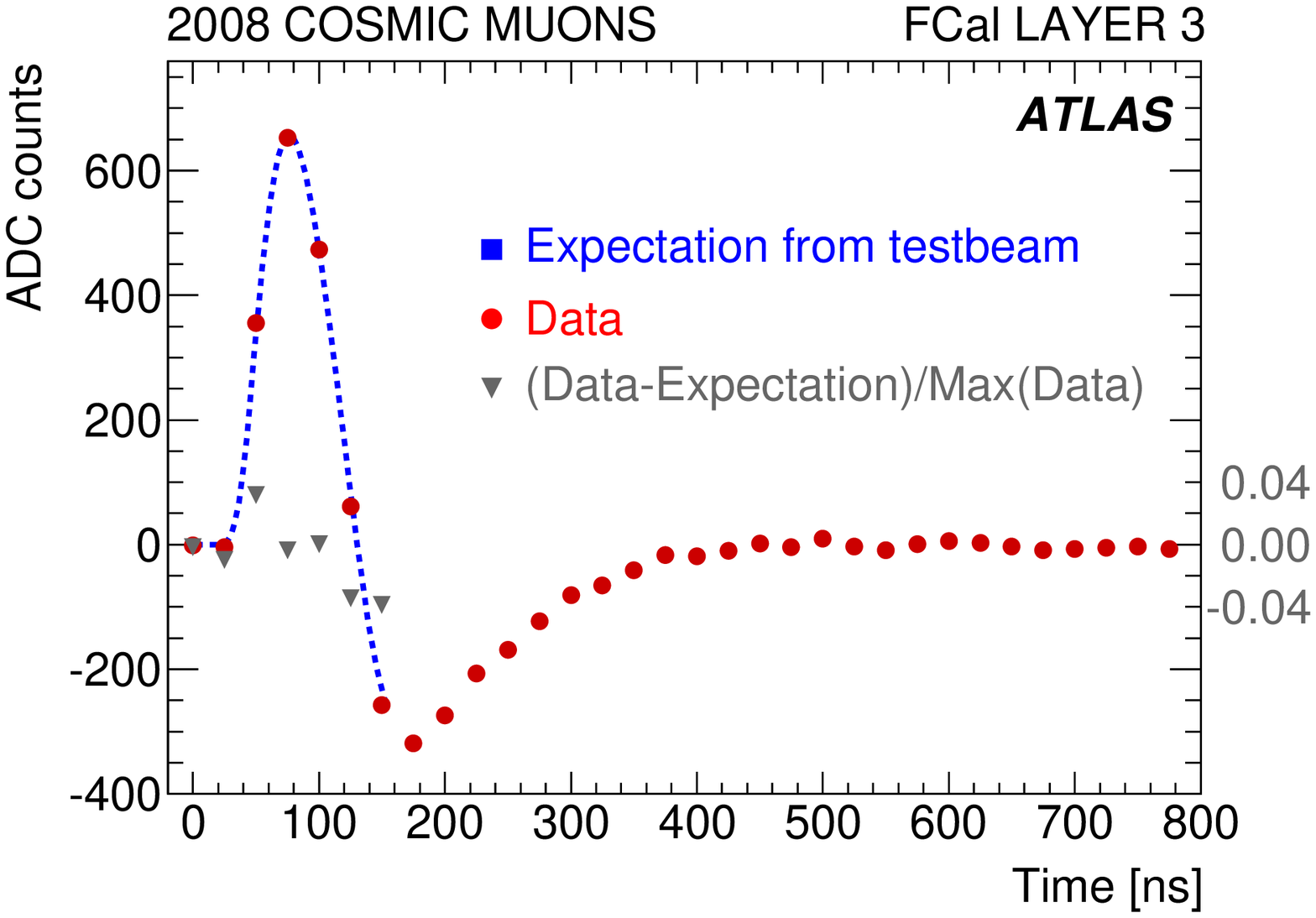,width=.49\linewidth, height=4.9cm} \\
\end{tabular}
\end{center}
\caption{\it Typical pulse shapes, recorded during the cosmic ray campaign, for a given cell in the second layer
  for the barrel (top left) and the endcap (top right) of the EM calorimeter,
  as well as in the first layer of the HEC (bottom left) and in the third layer of the FCal (bottom right). 
  The relative difference between
  data and prediction is indicated by triangles on the right scale.} 
\label{fig:pulse}
\end{figure*}

To illustrate the good quality of the pulse shape prediction, 
radiating cosmic muons depositing few GeV in a cell have been used. 
Figure~\ref{fig:pulse} shows a typical 32-sample pulse
recorded in the barrel (top left) and the endcap (top right) of the EM
calorimeter, as well as in the HEC (bottom left). In each case,
the pulse shape prediction, scaled to the measured cell
energy, agrees at the few percent level with the measured pulse.

As already mentioned, in the FCal the calibration pulse is injected at the base-plane 
of the front-end crates, and therefore the response to a calibration signal differs 
significantly from the response to an ionization pulse, preventing the use of methods 
described above. Instead, seven sample pulse shapes recorded during the beam test
campaign~\cite{ATLAS_TB_FCAL1,ATLAS_TB_FCAL2} have been 
averaged to obtain a normalized reference pulse shape for each layer. 
Figure~\ref{fig:pulse} (bottom right) shows a typical example where the
agreement between the reference pulse shape and the data is at the 4\% level.

\subsubsection{Quality of signal reconstruction in the EM calorimeter}
\label{sec:QualSigRec}

\begin{figure}[hbt]
\begin{center}
\includegraphics[width=0.45\textwidth]{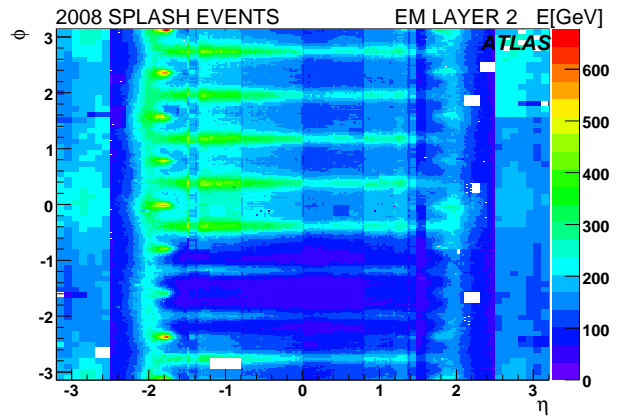}
\end{center}
\caption{\it Total energy deposited in the LHC beam splash events in
  every cell of the EM calorimeter second layer. Empty bins are due to
  non functioning electronics.}
\label{fig:Q1}
\end{figure}

Several PeV were deposited in the full calorimeter in LHC beam 
splash events. As an example, Figure~\ref{fig:Q1} shows the energy deposited 
in the second layer of the EM calorimeter. The structure in $\phi$ reflects
the material encountered by the particle flux before hitting the calorimeter, 
such as the endcap toroid. In this layer, a total of $5 \times 10^5$
five sample signal 
shapes with at least $5$~GeV of deposited energy were recorded. These
events were used to estimate the quality of the pulse shape prediction
for every cell.  

For this purpose, a $Q^2$-estimator is defined as~:  
\begin{equation}
Q^2 = \frac{1}{N_{\rm dof}} \sum_{j=1}^{N_{\rm samples}}
\frac{\left(s_{j}-A  g_{j}^{\rm phys}\right)^2}{\sigma_{\rm noise}^2+(kA)^2} , 
\label{eq:Q}
\end{equation}
where the amplitude $A$ (Eq.~(\ref{eq:Amax})) is computed with a number of samples $N_{\rm samples}=3$ (because 
the timing was not yet adjusted everywhere for the beam splash events, not all samples can be used),
$s_{j}$ is the amplitude of each sample $j$, in ADC counts, $g_{j}^{\rm phys}$ is the normalized predicted
ionization shape and $k$ is a factor quantifying the relative accuracy
of the amplitude $A$. Assuming an accuracy of around 1\%, with the $5$
GeV energy cut applied one has $\sigma_{\rm noise}^2 < (kA)^2$. In
this regime, it is possible to fit a $\chi^2$ function with 3 degree of freedom on
the $Q^2\times N_{\rm dof}$ distribution over cells in the central 
region (where the $Q^2$ variation is small). Therefore, $N_{\rm dof}=3$. A
given value of $Q^2$ can be interpreted as a precision on the amplitude at the
level $kQ$. 

\begin{figure}[hbtp]
\begin{center}
\includegraphics[width=0.45\textwidth]{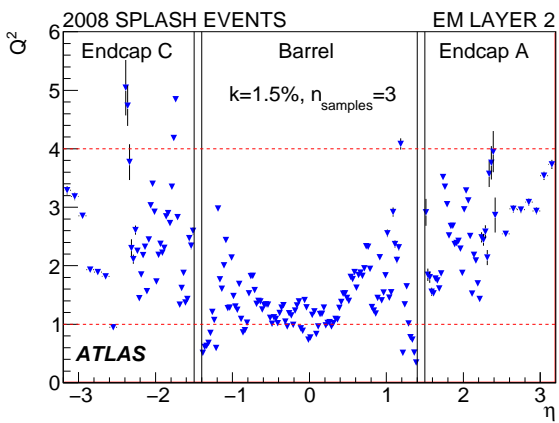}
\end{center}
\caption{\it Estimator $Q^2$ (defined in the
  text) as a function of $\eta$ for $5 \times 10^5$ pulse shapes 
  with $E>5$ GeV in the EM calorimeter second layer cells. $Q^2$ is
  defined in Eq.~(\ref{eq:Q}) with $k=1.5\%$.}
\label{fig:Q2}
\end{figure}

Figure~\ref{fig:Q2} shows the $Q^2$-estimator in the second
layer of the EM calorimeter averaged over $\phi$, assuming $k=1.5\%$
corresponding to $Q^2\sim1$ for $\eta\sim0$. The accuracy is degraded
by at most a factor of $\sim$2 (i.e. $Q^2 \sim 4$) in some endcap
regions. This shows that these data can be described with a reasonable
precision.  

\subsubsection{Ionization electron drift time measurement in the EM calorimeter}
\label{sec:tdrift}

During the 2008 cosmic runs, half a
million pulses with 32 samples were recorded in the EM calorimeter from cells in which at
least 1 GeV was reconstructed. Given the good
accuracy of the predicted signal undershoot (see Figure~\ref{fig:pulse}), the drift time can be 
extracted  from a fit to the measured signal~\cite{TDRIFT2}.

\begin{figure}[hbtp]
\begin{center}
\includegraphics[width=0.45\textwidth]{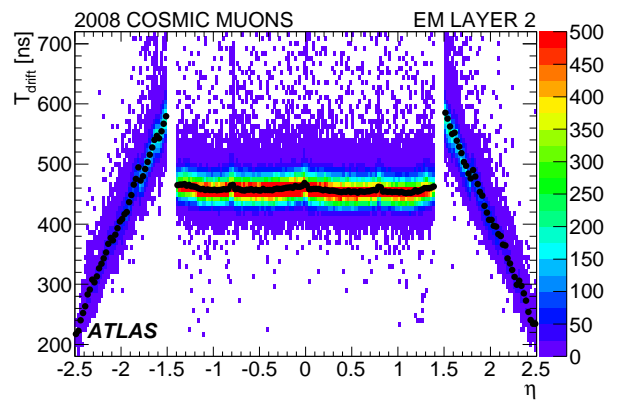}
\end{center}
\caption{\it Drift time measurement in the cells of the EM
  calorimeter second layer with $E>1$~GeV for the 2008 cosmic muon
  run. The dots correspond to drift time values averaged in
  $\phi$.}
\label{fig:tdrift1}
\end{figure}

Figure~\ref{fig:tdrift1} shows the fitted drift time for all 
selected cells in the second layer using the standard pulse shape
prediction method (Section~\ref{sec:PredSigRec}).  
In the EMB, the drift time 
has also been measured with a method in which the shape is computed using 
a more analytical model and $LC$ and $RC$ extracted from network analyzer measurements~\cite{SHAPE0}. 
The drift times extracted from the two methods
are in excellent agreement, giving confidence in the results: a constant value around
the expected 460~ns is obtained, except near the electrode
edges ($|\eta|=0, 0.8$ and $1.4$) where the electric field
is lower. The decrease of the drift time in the EM endcap
($1.5<|\eta|<2.5$) reflects the decrease of the gap size with
$|\eta|$. Similar results are obtained for the first and third layers of
the EM calorimeters. 

\subsubsection{Impact on the global energy resolution constant term of
the EM calorimeter}
\label{sec:CT}

When five of the production EM calorimeter modules were tested
individually in electron beams,  
the global constant term $c$ of the energy resolution
formula was measured to be $c\sim0.5\%$ in the EM barrel and 0.7\% in the EM endcap~\cite{ATLAS_TB_EM3}. 
The main contributors are the signal reconstruction accuracy, the
LAr gap uniformity, 
and the electronics calibration system. The first two
contributions $c_{\rm SR}$ and $c_{\rm gap}$ can be investigated using results
presented in Section~\ref{sec:QualSigRec} 
and~\ref{sec:tdrift}, considering only the second layer of the EM
calorimeter where most of the electromagnetic shower energy is
deposited. 

From Figure~\ref{fig:Q2} , one finds that $<Q^2> \sim 1.4$ in the EM barrel and $2.6$ in
endcap, and hence $<k>=1.8\%$ and $2.4\%$ respectively. This
corresponds to residuals between the predicted and measured
pulses of 1 to 2\% of the pulse amplitude (see Figure~\ref{fig:pulse}
for illustration), for samples around the signal maximum.  
Similar residuals were obtained in the electron beam test
analysis~\cite{RTM}. At this time, the contribution of the signal
reconstruction to the constant term was estimated to be $c_{\rm 
  SR}=0.25\%$. Given the measured accuracy with beam splash
events, the beam test result seems to be reachable 
with LHC collisions. 

\begin{figure}[hbtp]
\begin{center}
\includegraphics[width=0.45\textwidth]{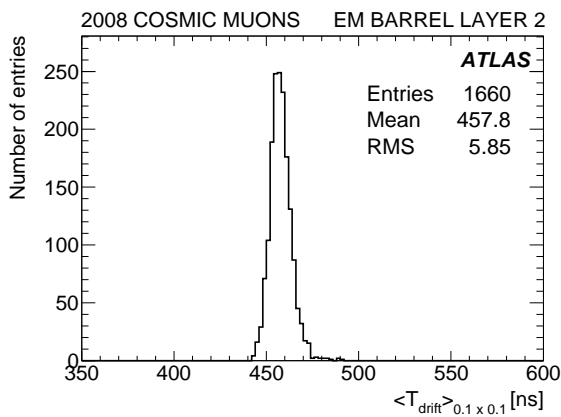}
\end{center}
\caption{\it Distribution of the local average drift time values
  in $\Delta\eta \times \Delta\phi= 0.1 \times 0.1$ bins, for the middle layer of the
  EM barrel.} 
\label{fig:tdrift2}
\end{figure}

The drift time measured in Section~\ref{sec:tdrift} is a function of the
gap thickness ($w_{\rm gap}$) and the high voltage ($V$):
\begin{equation}
T_{\rm drift} \sim \frac{w_{\rm gap}^{\alpha+1}}{V^{\alpha}}
\label{eq:tdrift}
\end{equation}
where $\alpha\simeq 0.3$ is empirically determined from
measurements~\cite{TDRIFT}. In the EM barrel, the electric field is
constant, except in transition regions, and thus the drift time 
uniformity directly measures the LAr gap variations. To reduce statistical
fluctuations, the measured drift time values are averaged over regions
of $\Delta \eta
\times \Delta \phi= 0.1 \times 0.1$. The distribution of the average drift
time is shown in Figure~\ref{fig:tdrift2} for the second layer
of the EM barrel calorimeter. 

The drift time uniformity, estimated as the ratio of the RMS of 
this distribution to its mean value, is $1.28\pm0.03\%$.  
Using the relation between the drift time and the gap from 
Eq.~\ref{eq:tdrift} and the fact that the signal amplitude
is proportional to the initial ionization current 
($I \simeq \frac{\rho \cdot w_{\rm gap}}{T_{\rm drift}} \simeq w_{\rm gap}^{-\alpha}$ 
where $\rho$ is the linear density of charge), one can relate the
relative variation of the drift time to the one of the amplitude
applying a factor $\alpha/(1+\alpha)$ to the above result. 
Therefore, the drift time uniformity leads to a dispersion 
of response due to the barrel calorimeter gap variations of 
$(0.29^{+0.05}_{-0.04})\%$ where the systematic uncertainties
are included. 
This represents an upper bound on the corresponding 
constant term $c_{\rm gap}$.


For comparison, during the EM calorimeter barrel module
construction, the LAr gap thickness was measured, yielding
an estimate of the constant term due to gap size variations of
$c_{\rm gap}= 0.16\%$~\cite{EMB_CONS}. 
The measurement of the gap size  uniformity presented  here  takes
into account further  effects like deformations in the assembled
wheels and possible systematic uncertainties from the in situ cosmic
muon analysis.

%% file: Uniformity_epj.tex
\section{In situ EM calorimeter performance with cosmic muons}
\label{sec:analysis}

In the previous sections, we demonstrated the good performance of
the electronics operation and the good understanding of
the energy reconstruction. The cosmic ray events can therefore now be used
to validate the Monte Carlo simulation that will be used for the first
collisions. 

Two such analyses are presented in this section: the first study aims
to investigate the electromagnetic barrel calorimeter uniformity using  
ionization signals from quasi-projective cosmic muons, and the second 
aims to reconstruct electromagnetic showers from radiative cosmic muons
and to compare the measured shower shapes with simulation. 

\subsection{Monte Carlo simulation}

The ATLAS Monte Carlo~\cite{MCATLAS,CALOMC} simulates the interaction of
particles produced during LHC collisions or from cosmic muons
within the ATLAS sub-detectors. It is based on the
Geant4 toolkit~\cite{geant4} that provides the physics lists, geometry
description and tracking tools. For cosmic muons, the material between
the ground level and the ATLAS cavern is also simulated, i.e.
the overburden and the two access shafts. The simulated cosmic ray
spectrum corresponds to what was measured at
sea level~\cite{cosmics}. Air showers are not simulated but have a
negligible effect on the analyses presented here. 
In order to save CPU time, the generated
events are filtered before entering the full Geant4 simulation by
requiring that the particles cross a specific detector volume (in the
following analyses, typically inner detector volumes). 

An important use of the simulation, 
amongst many others, is to validate the selection criteria on shower-shape 
for high-level trigger and offline algorithms, as well as to derive
the electron and photon energy calibrations. 

It is important to note that, thanks to the digitization step of the
calorimeter simulation which emulates the behavior of the
electronics, the standard energy reconstruction procedure can be
applied to the simulated events. The special procedure used for
asynchronous cosmic muon data, which uses an iterative determination of
the event time, is however not applied to the Monte Carlo data.  

\subsection{Uniformity of the electromagnetic barrel calorimeter}
\label{sec:uniformity}

\subsubsection{Goals and means of the analysis}

Any non-uniformity in the response of the calorimeter has a direct
impact on the constant term in the energy resolution (see
Section~\ref{sec:CT}); great care was taken during the construction to
limit all sources of non-uniformity to the minimum achievable, 
aiming for a global constant term 
below $0.7\%$. The default ATLAS Monte Carlo simulation 
emulates the effect of the constant term, but for the present
analysis, this emulation was turned off. 

The uniformity of the calorimeter was measured for three barrel
production modules using electrons during beam test
campaigns~\cite{ATLAS_TB_EM3}. Cosmic muons 
provide a unique opportunity to measure the calorimeter uniformity 
in situ over a larger number of modules, unfortunately limited to the
barrel calorimeter due to both the topology of the cosmic muon events  
and the choice of triggers. The scope of this analysis is nevertheless quite
different than in the beam test. First, muons behave very differently from
electrons: in most events, they deposit only a minimum ionization
energy in the liquid argon and they are much less sensitive to upstream
material. The result can therefore not be easily extrapolated to the
electron and photon response. Second, the cosmic run statistics are
limited, so uniformity cannot be studied with cell-level granularity. 
The goal of this cosmic muon
analysis is rather to quantify the agreement between data and Monte
Carlo, and to exclude the presence of any significant non-uniformity in the
calorimeter response.  

A previous uniformity analysis using cosmic muons~\cite{ATLAS_UNIF1}
from 2006 and 2007 relied on the hadronic Tile calorimeter to trigger
events and to measure the muon sample purity. For the 2008 data
discussed here, both the muon spectrometer and inner detector
were operating and were used for triggering and event selection. The
data sample consists of filtered events requiring a reconstructed
track in the inner detector with at least one hit in the silicon
tracker. The tracks are also selected to be reasonably projective by
requiring that their transverse ($|d_{\rm 0}|$) and longitudinal
($|z_{\rm 0}|$) impact parameters, with respect to the 
center of the coordinate system be smaller than 300~mm.   

\subsubsection{Signal reconstruction}

In the first step, a muon track is reconstructed in the inner
detector. For that purpose, a dedicated algorithm looks for a single
track crossing both the top and bottom hemispheres. This single
track is then extrapolated both downward and upward into the calorimeter.    

Around the two track impact positions in the calorimeter, a rectangle
of cells (the cell road) is selected in the first and second layers (the
signal to noise ratio for muons is too low in the third layer). The
cells of the first layer have a size of 
$\Delta\eta \times \Delta\phi = 0.003 \times 0.1$ and $12\times 3$ such cells are
kept. Similarly, the cells of the second layer have a
size of $\Delta\eta \times \Delta\phi = 0.025\times0.025$, and $5\times 5$
such cells are kept.   

To reconstruct the energy of the selected cells, the muon timing is obtained
via an iterative procedure that is usually only applied to cells with
an ADC signal at least four times the noise level. Since most
muons are minimum ionizing particles, the muon signal is small,
typically $150$ MeV is deposited in the most energetic cell in the
second layer, only five times the noise, and many cells do not pass this
threshold. Therefore, an alternative reconstruction is used in this
analysis: in the first pass, the iteration threshold is lowered to zero
so that the timing is computed for most of the cells. In the second
pass, the timing of the most energetic cell determined in the first
pass is applied to all the other cells of the road. The cell
energy is reconstructed at the electron energy scale and thus does
not represent the true energy loss of the muon. Finally, clusters are
formed in each layer to reconstruct the muon energy loss. The criteria
used to decide on the cluster size are described below. 

\subsubsection{Optimization of the uniformity measurement}

In order to perform the most accurate evaluation of the calorimeter
uniformity, the measurement granularity, the cluster size and the
selection cuts have been optimized. The granularity chosen
is a compromise between the need for high statistics (large
binning) and the need for high precision. The cluster size optimizes
the signal to noise ratio while the selection cuts
reduce the biases while keeping high statistics.

The binning is determined by requiring a minimum of $500$
events per unit. In the $\eta$ direction, this corresponds to bins of
$0.025$ (equal to the second layer cell width) up to $|\eta|=0.7$ and
wider bins above.  

In the first layer, the muon energy loss is measured using a $\Delta
\eta \times \Delta \phi = 2\times 1$ (in first layer cell unit) cluster, which
contains most of the deposited energy. Adding an 
additional cell brings more noise than signal. In the second
layer, a $1\times 3$ (in second layer cell unit) cluster is used:
it suffers less from noise than 
a $3\times 3$ cluster, but requires
the removal of non-projective 
events which leak outside the cluster along the $\eta$ direction. 

This projectivity cut is based on the centrality of the muon in the
second layer cell: when the muon passes close to the edge of the
cell, a very small non-projectivity induces a large energy leakage
into the neighboring cell. Therefore, for each second layer cell, eight bins
corresponding to the eight first layer cells located in front
of it
were defined, and in each bin a cut is applied on the beam
impact parameter $z_{\rm 0}$ of the track, such that the muon is
geometrically contained in the second layer cell. The remaining statistics
after this projectivity cut is 76~k events in the data sample and 113~k
events in the Monte Carlo sample. The events are mainly located under
the cavern shafts leading to a coverage of
around $20\%$ of the full electromagnetic barrel calorimeter. 

\begin{figure}[hbtp]
\begin{center}
\includegraphics[width=0.45\textwidth]{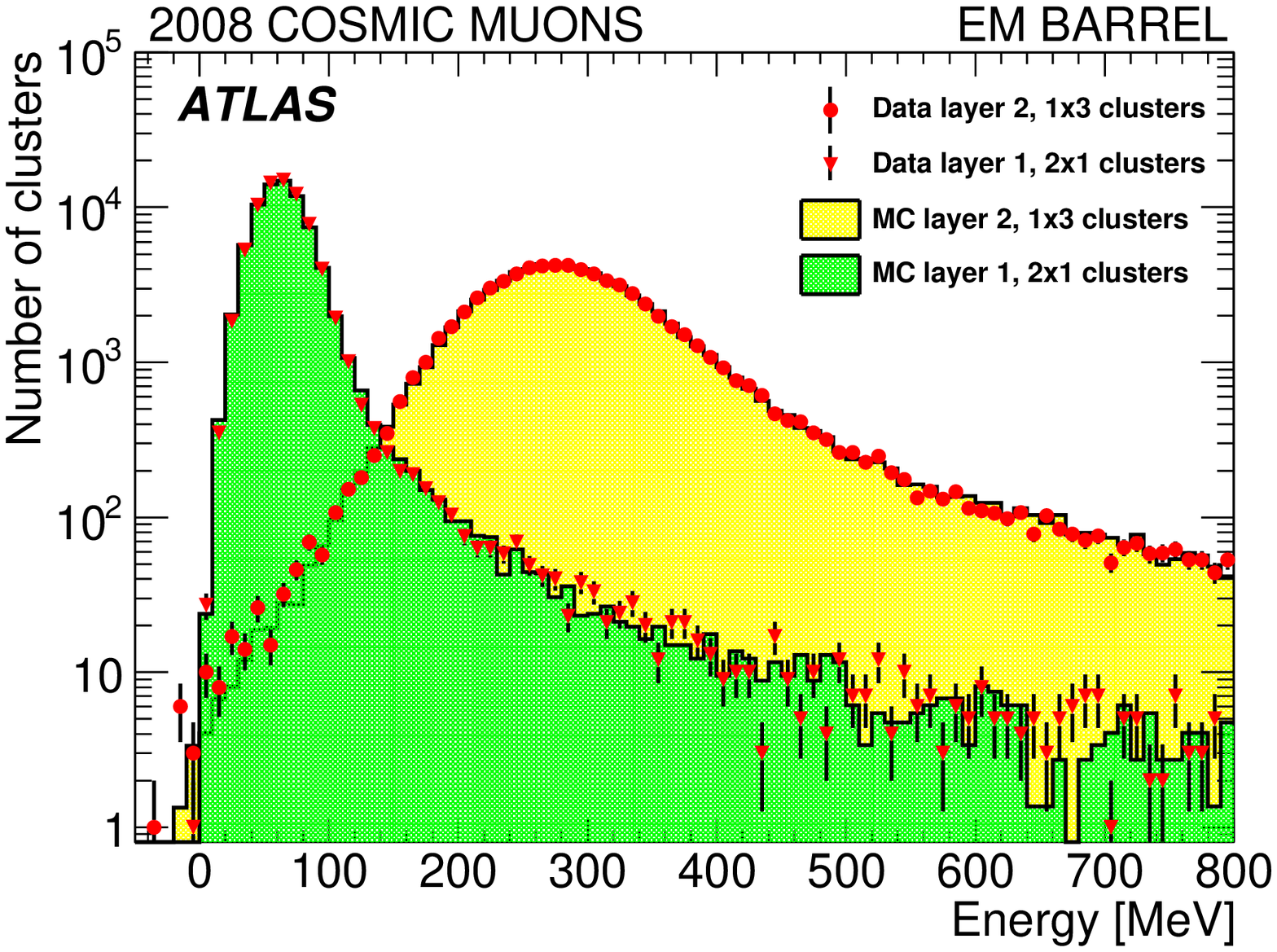}
\includegraphics[width=0.45\textwidth]{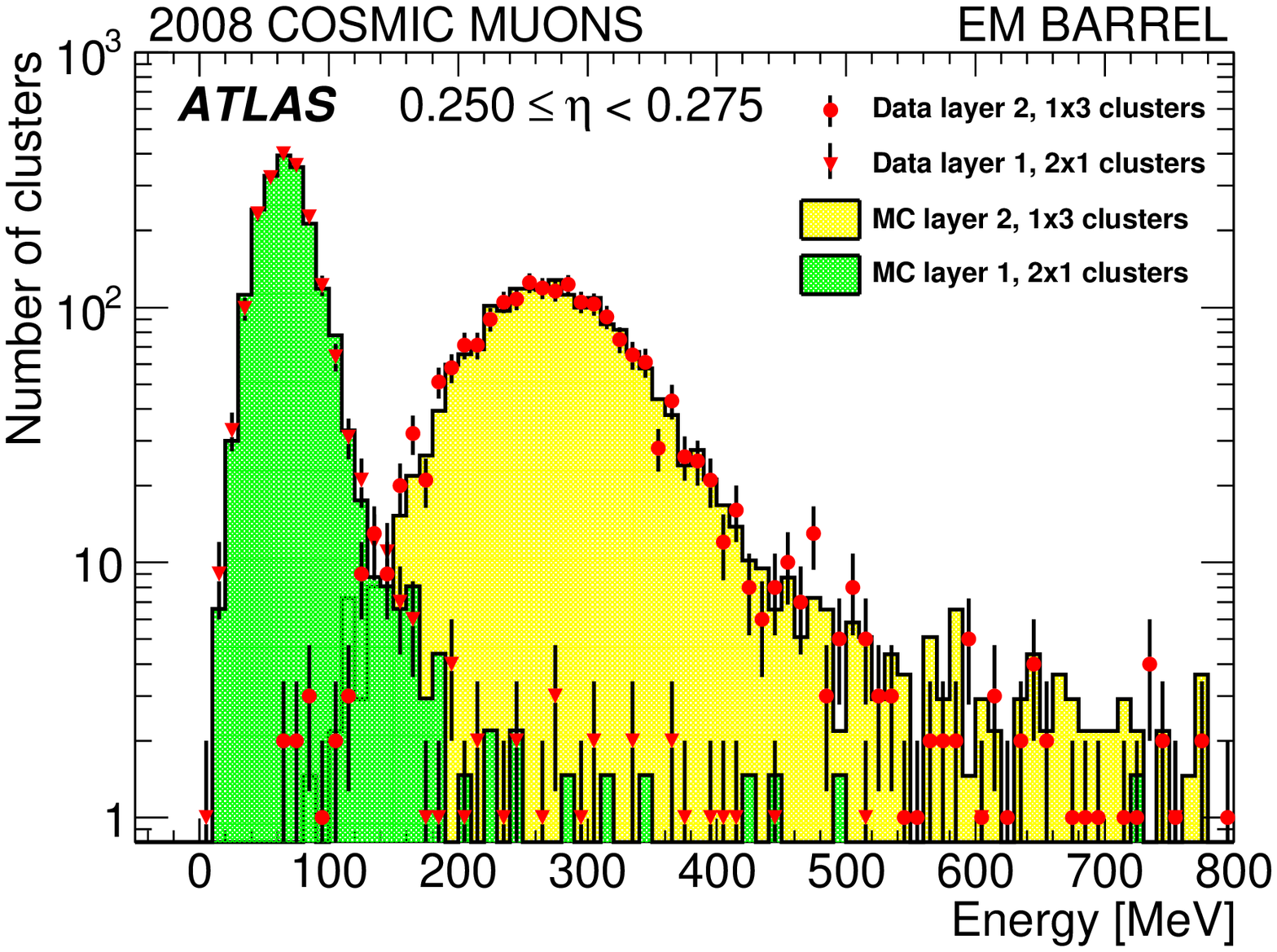}
\end{center}
\caption{\it  Energy in a $2\times 1$ cluster in the first layer 
  (histogram for Monte Carlo and triangles for data) and in a
  $1\times 3$ cluster in the second layer (histogram for
  Monte Carlo and full circles for data) for all events (top) and a
  single $\eta$-bin (bottom).  }
\label{fig:lineshape}
\end{figure}

A comparison of the energy reconstructed in the first and second
layers between data and Monte Carlo events is shown in
Figure~\ref{fig:lineshape}. Because the muon energy loss is mostly
$\eta$-dependent, both distributions are shown for all events
(top), showing a large width due to the variation of the energy
response over $\eta$, and for a single $\eta$-bin (bottom). 

The agreement between the data and Monte Carlo distributions is very
good, both for the shape and for the absolute energy scale which
differs by only $2\%$ in the front layer and $1\%$ in the second
layer. Part of the difference comes from the slight difference in 
acceptance for data and Monte Carlo, as well as from the difference in 
energy reconstruction. 
This overall energy scale difference is
corrected for in the MC in the rest of the study.  

\subsubsection{Calorimeter uniformity along $\eta$}

Given the limited statistics of the projective cosmic muon data, the
uniformity of the response in $\eta$ cannot be estimated at the cell
level. A natural choice of cell combination is to integrate clusters
in $\phi$ since the response should not vary along this direction due
to the $\phi$ symmetry of the calorimeter. The response along the
$\eta$ direction for cosmic muons depends on the 
variation of the amount of liquid argon seen by the muon. In
particular, a transition occurs at $|\eta|=0.8$ where the lead 
thickness goes from $1.53$~mm to $1.13$~mm. 

The estimation of the muon energy in each $\eta$-bin is done with a
fit of the cluster energy distribution using a Landau function convoluted
with a Gaussian. The Landau function accounts for fluctuations of
the energy deposition in the ionization process and the Gaussian
accounts for the electronic noise and possible remaining fluctuations.
In particular, a $10\%$ difference is observed between the width of the Gaussian expected
from the electronic noise and the width of the fitted Gaussian. 
Mostly this bias comes from remaining cluster non-containment
effects which are found to be $\eta$-independent and thus do not produce any
artificial non-uniformity. The most probable value (MPV) of the
Landau distribution estimates the energy deposition.  

\begin{figure}[hbtp]
\begin{center}
\includegraphics[width=0.5\textwidth]{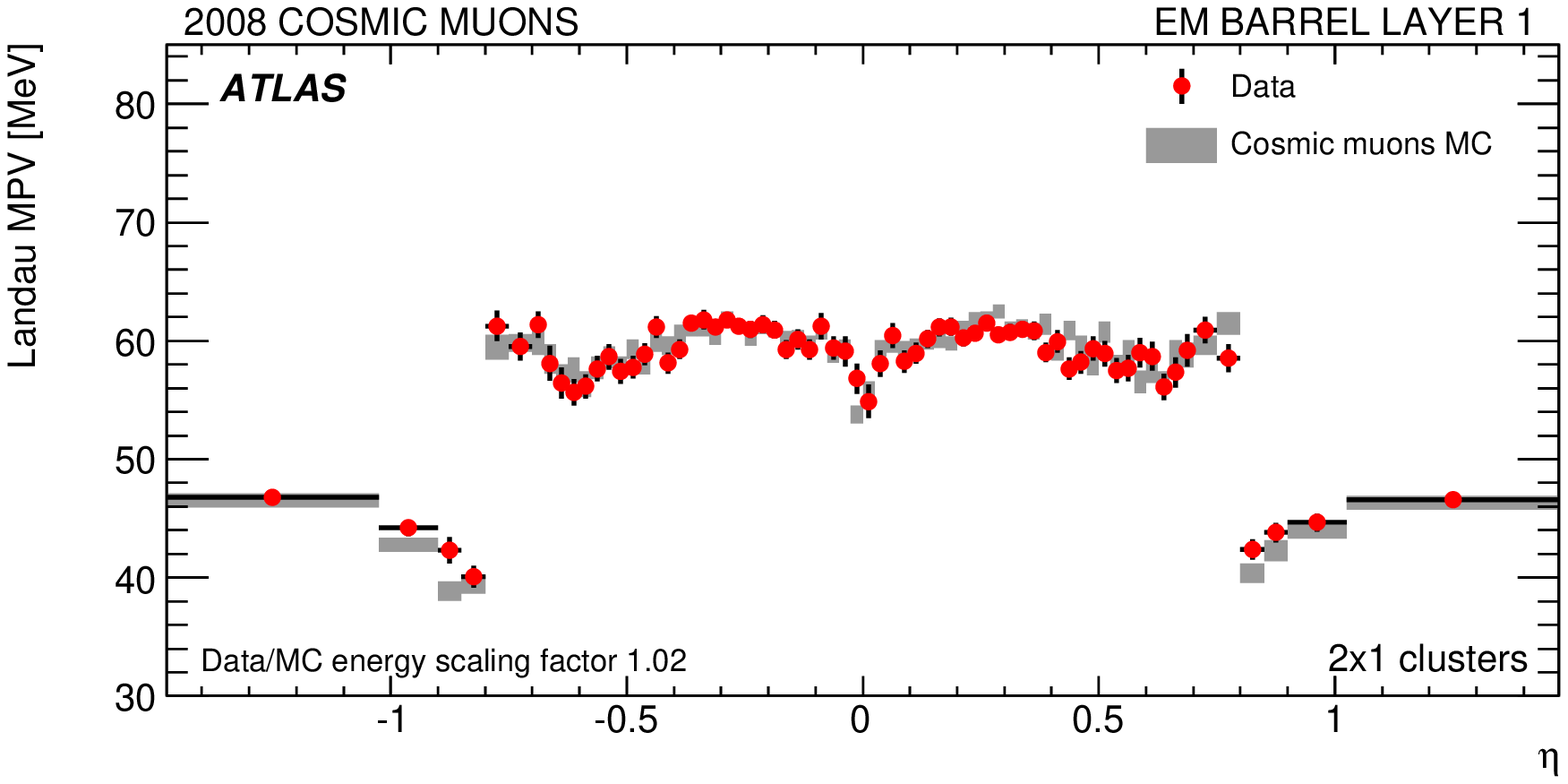}
\includegraphics[width=0.5\textwidth]{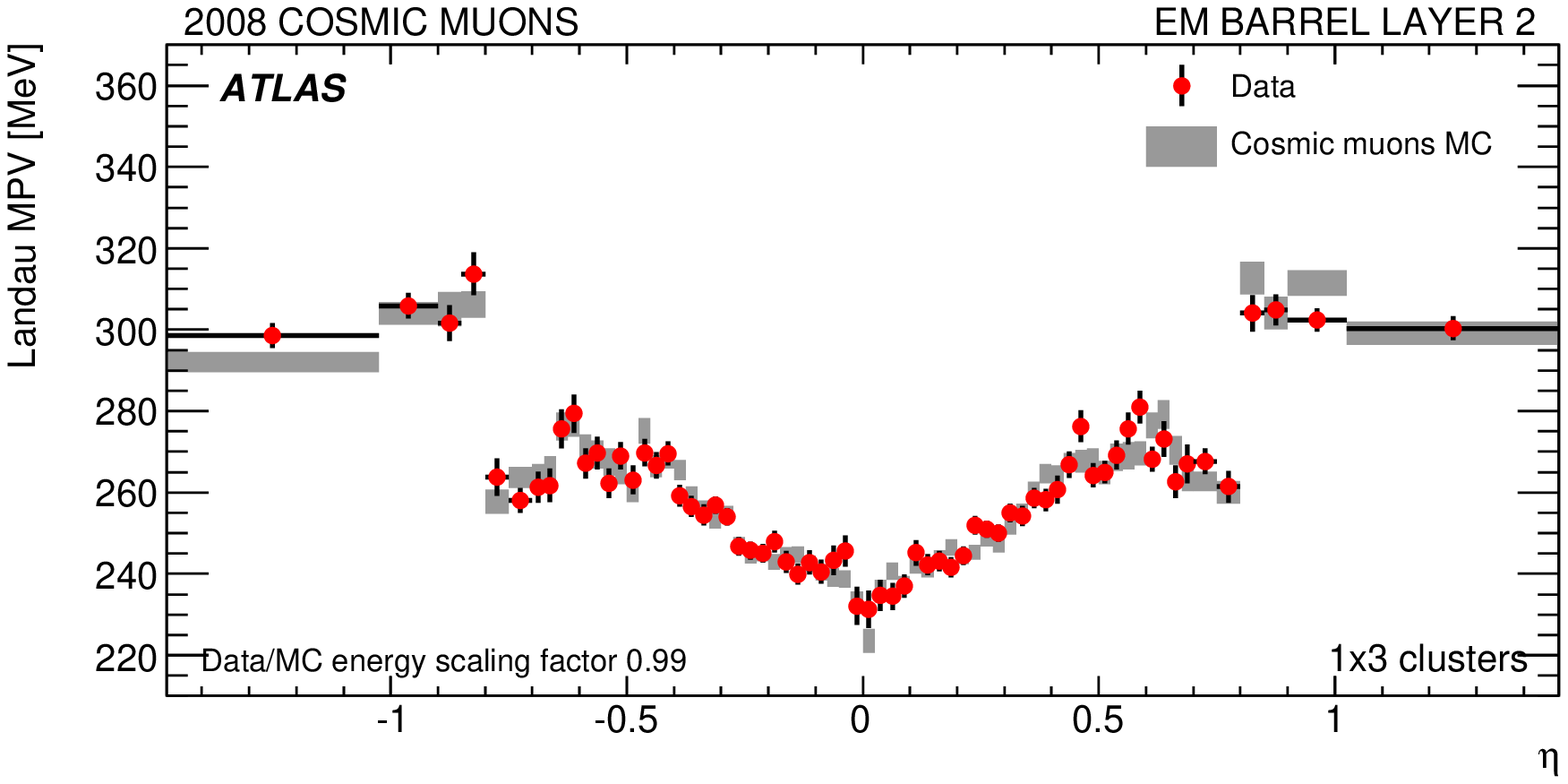}
\caption{\em Landau MPV as a function of $\eta$ in the first (top) and
  second (bottom) layers for the data (red points) and Monte Carlo
  (grey bands). }
\label{fig:MPV}
\end{center}
\end{figure}

Distributions of data and Monte Carlo MPVs along the $\eta$
direction for the first and second layers are shown in
Figure~\ref{fig:MPV}. 

In the first layer, the MPVs are roughly constant along $\eta$, except
around $\eta=0$ where some cells are physically missing in the
detector, and around $|\eta|=0.6$ where the cell depth is varying. 
In the second layer, the response follows a typical
``V-shape'' corresponding to the variation of the cell depth along
$\eta$ that rises up to $|\eta|=0.6$. Again, the agreement between the
data and Monte Carlo is very good, showing that the contribution of
systematic effects due to the energy reconstruction method
or the non-projectivity of the tracks is small.

The response uniformity $U_{\rm meas}$ is given by the RMS of the normalized differences
between the data and Monte Carlo MPVs in each $\eta$-bin
:\begin{equation} 
\label{eq:measU}
U_{\rm meas} =  \sqrt{\frac{\sum_{i=1}^{
      N_{\rm b}}{(U_{i, {\rm meas}}- < U_{i, {\rm meas}}>)^2}}{N_{\rm b}}},
\end{equation}
with:
\begin{equation} 
U_{i, {\rm meas}} = \frac{MPV_{i, {\rm Data}}-MPV_{i, {\rm MC}}}{MPV_{i,
  {\rm Data}}}~,
\end{equation}
where $U_{i, {\rm meas}}$ is averaged over $\phi$, 
$N_{\rm b}$ is the number of bins in $\eta$, and $<$$U_{i, {\rm meas}}$$>$=0 
since the global energy scale difference was corrected by rescaling the MC.

The measured uniformity should be compared to the expected uniformity $U_{\rm exp}$, which is obtained similarly to Eq.~\ref{eq:measU} with $U_{\rm i, exp}$ given by:
\begin{equation}
\label{eq:uiexp}
U_{i, {\rm exp}} = \frac{MPV_{i,{\rm MC}}}{MPV_{i,{\rm Data}}}
\sqrt{U_{i,{\rm Data}}^2+U_{i,{\rm MC}}^2}
\end{equation}
with:
\begin{equation}
U_{i, {\rm Data(MC)}} = \frac{\sigma(MPV_{i,{\rm
            Data(MC)}})}{MPV_{i,{\rm Data(MC)}}},
\end{equation}
where $\sigma(MPV_{i,{\rm Data(MC)}})$ is the statistical uncertainty on the
measured Landau MPV. This uncertainty is due to the finite statistics
of the data and Monte Carlo samples in each bin, the Landau dispersion of the
ionization, and the electronic noise. 

The measured uniformity $U_{\rm meas}$ should agree with the expected
uniformity $U_{\rm exp}$ if the Monte Carlo simulation reproduces 
the data well: the key ingredients are the acceptance, the muon spectrum, and the
energy reconstruction method. A significant departure of the measured
uniformity from the expected one would be a measurement of
additional non-uniformities $U_{\rm \Delta}$ ($U_{\rm
  \Delta}^2=U_{\rm meas}^2-U_{\rm exp}^2$). 

\begin{figure}[hbtp]
\begin{center}
\includegraphics[width=0.5\textwidth]{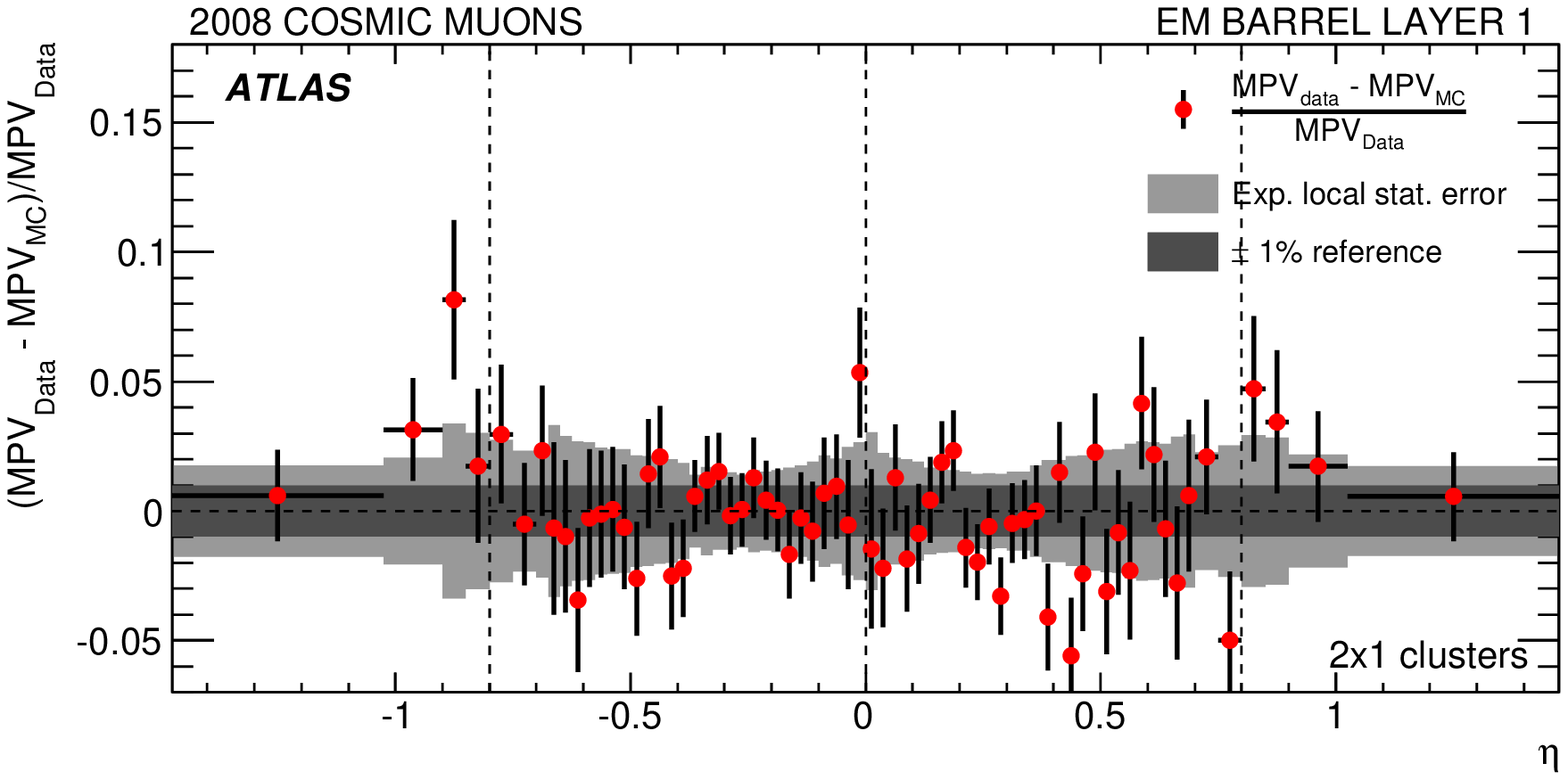}
\includegraphics[width=0.5\textwidth]{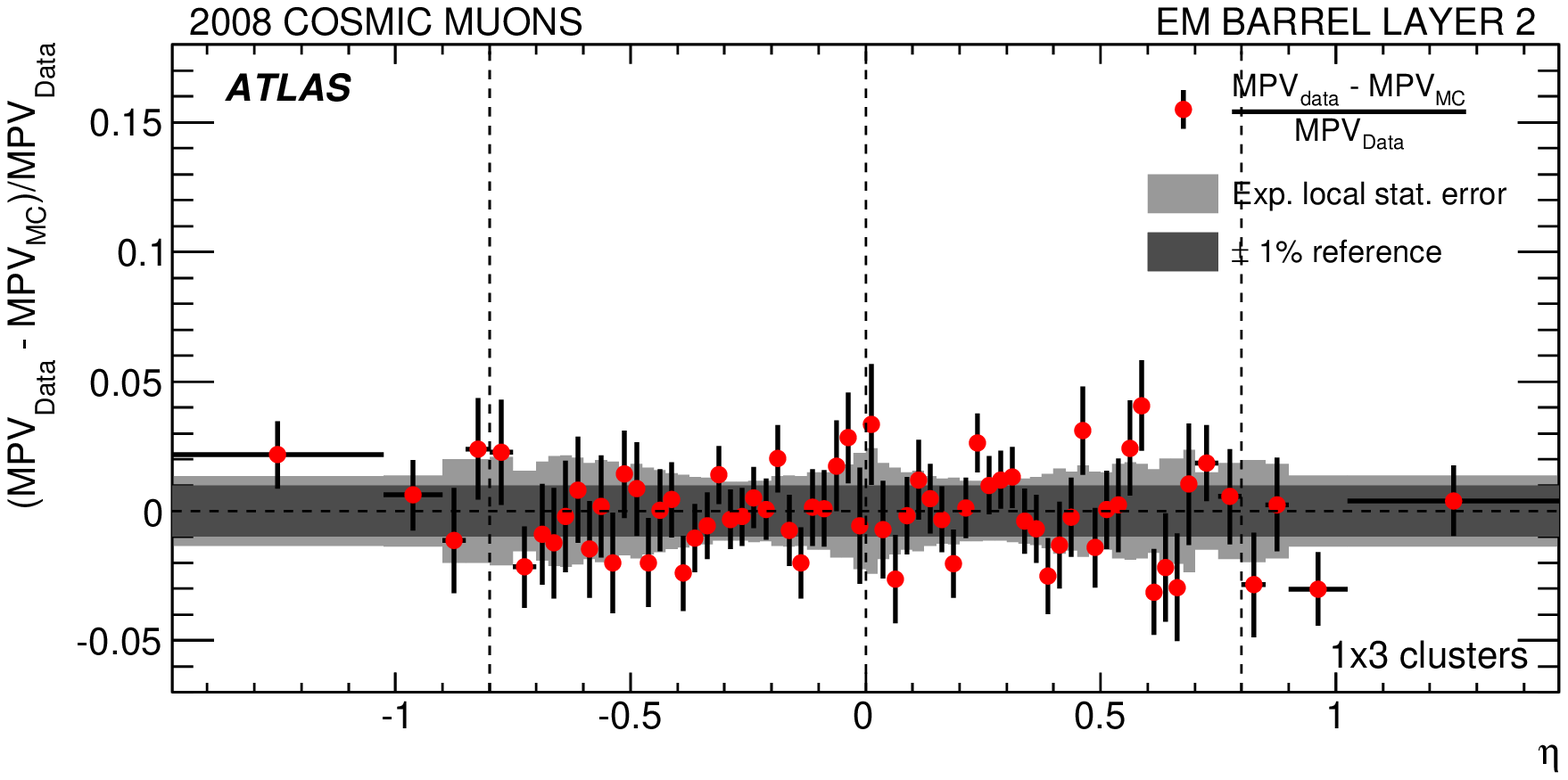}
\caption{\em Measured $U_{i,{\rm meas}}$ (red points) and expected
  $U_{i,{\rm exp}}$ (light grey band) cosmic muon energy dispersions
  as function of $\eta$ for the first (top) and second (bottom)
  layers of the EM barrel. The dark grey band indicates a $\pm 1\%$ strip for reference.}
\label{fig:unifL}
\end{center}
\end{figure}

The measured and expected uniformities for the two EM layers are
shown in Figure~\ref{fig:unifL}. 

The fluctuations of the measured energies are large: the RMS of
the corresponding distribution is $2.4 \pm 0.2 \%$ in the first layer and $1.7
\pm 0.1 \%$ in the second layer, showing that the statistical power
of the analysis is limited given the available data and Monte Carlo
statistics. The fluctuations mostly remain within the limits of
the band representing the expected values. The
RMS of the latter distribution is 2.2 $\%$ in the first
layer and 1.6 $\%$ in the second layer. This demonstrates that no significant
additional non-uniformity ($U_{\rm \Delta}$) is present in the data. An upper
limit is derived and yields $U_{\rm 
  \Delta} < 1.7\%~@~95\%~{\rm CL}$ in the first layer, and $U_{\rm \Delta} <
1.1\%~@~95\%~{\rm CL}$ in the second layer.      

The calorimeter response uniformity along $\eta$ (averaged over
$\phi$) is thus consistent at the percent level with the
Monte Carlo simulation and shows no significant non-uniformity. 

%% file: Photons_epj.tex
\subsection{Electromagnetic shower studies}

The second analysis aims at validating the Monte Carlo simulation of
the distribution of some key calorimeter variables 
used in the ATLAS electron/photon identification. This is done using
radiative cosmic muons that can give rise to electromagnetic showers in the
calorimeter through bremsstrahlung or pair conversions. 

\subsubsection{Selection of radiative muons}

To increase the probability of the presence of a muon in the event, it is requested that
at least one track has been reconstructed in the inner detector
barrel with $|d_{\rm 0}|<220$ mm and $p_{\rm T}>5$ GeV: these cuts ensure 
a similar acceptance for data and Monte Carlo. 

\begin{figure}[hbtp]
\begin{center}
\includegraphics[width=0.45\textwidth]{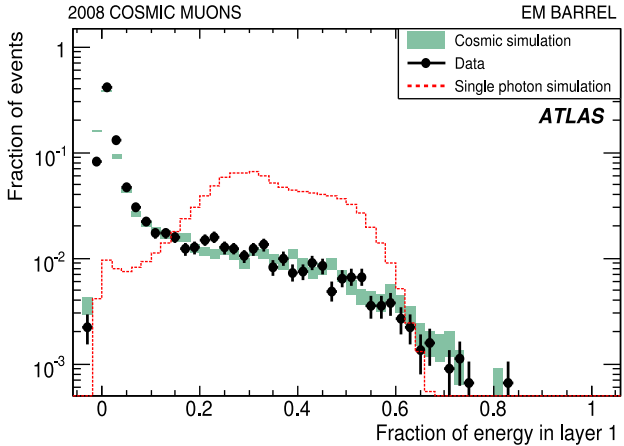}
\end{center}
\caption{\it Fraction of cluster energy deposited
  in the first layer of the electromagnetic barrel calorimeter for cosmic
  data (dots) and Monte Carlo (rectangles), as well as for
  simulated single photons of $5$ GeV momentum from interaction vertex
  (red histogram).} 
\label{fig:photons1a}
\end{figure}

A radiative energy loss is searched for in the electromagnetic barrel
calorimeter by requiring a cluster with an energy greater than $5$ GeV. Since the
radiation can occur anywhere along the muon path, the corresponding
shower is not always fully contained in the electromagnetic
calorimeter: this is visible in Figure~\ref{fig:photons1a} which
shows the fraction of the cluster energy deposited in the first layer
for simulated single photons from interaction vertex and for electromagnetic showers from radiating
cosmic muons. This shows that the longitudinal shower
development of the radiative photons is well reproduced by the
Monte Carlo simulation, and that most of the radiating muons
deposit very little energy in the first layer. To select ``collision-like'' 
showers, this fraction is requested to be greater than $0.1$. A total of $1200$ 
candidates remain in the data sample and $2161$ in the Monte Carlo after this
selection.     

\subsubsection{Shower shape validation}

\begin{figure}[hbtp]
\begin{center}
\includegraphics[width=0.45\textwidth]{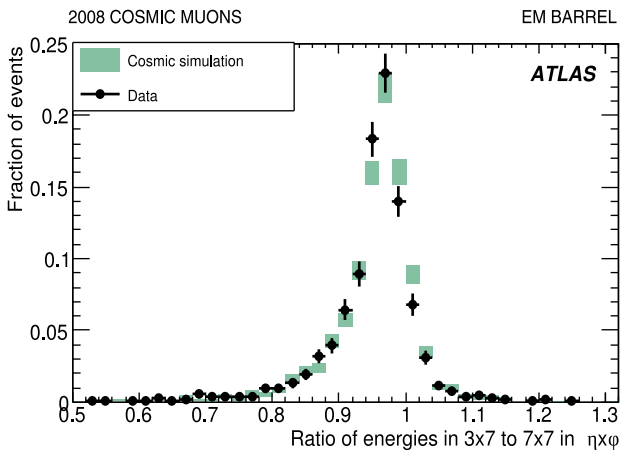}
\end{center}
\caption{\it Lateral
  shower containment in the second layer of the calorimeter given by
  the ratio of the energy deposited in a $3\times 7$ cluster to a
  $7\times 7$ cluster for radiative cosmic muon data (dots) and Monte
  Carlo simulation (rectangles).} 
\label{fig:photons1b}
\end{figure}

\begin{figure}[hbtp]
\begin{center}
\includegraphics[width=0.42\textwidth]{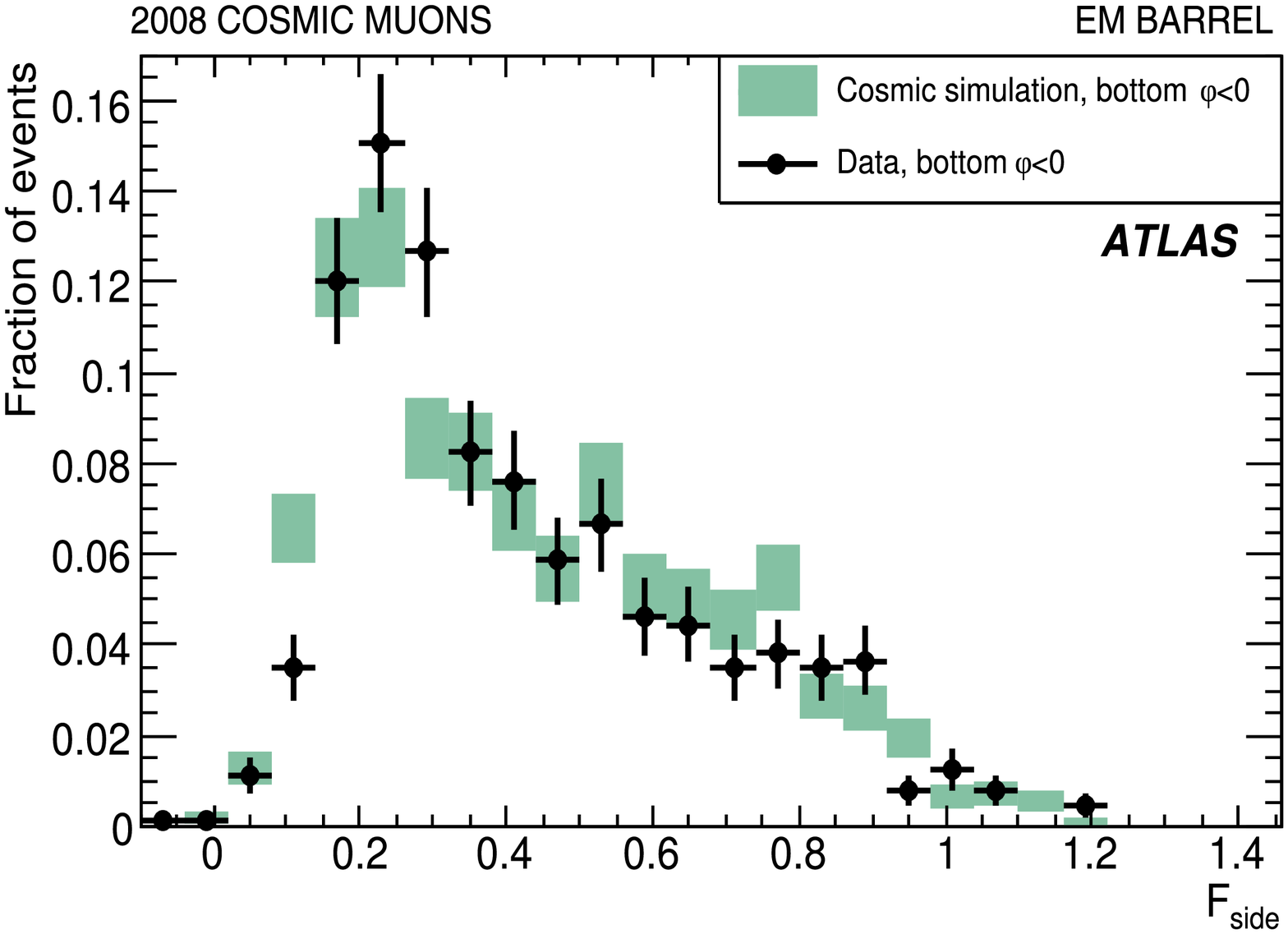}
\includegraphics[width=0.42\textwidth]{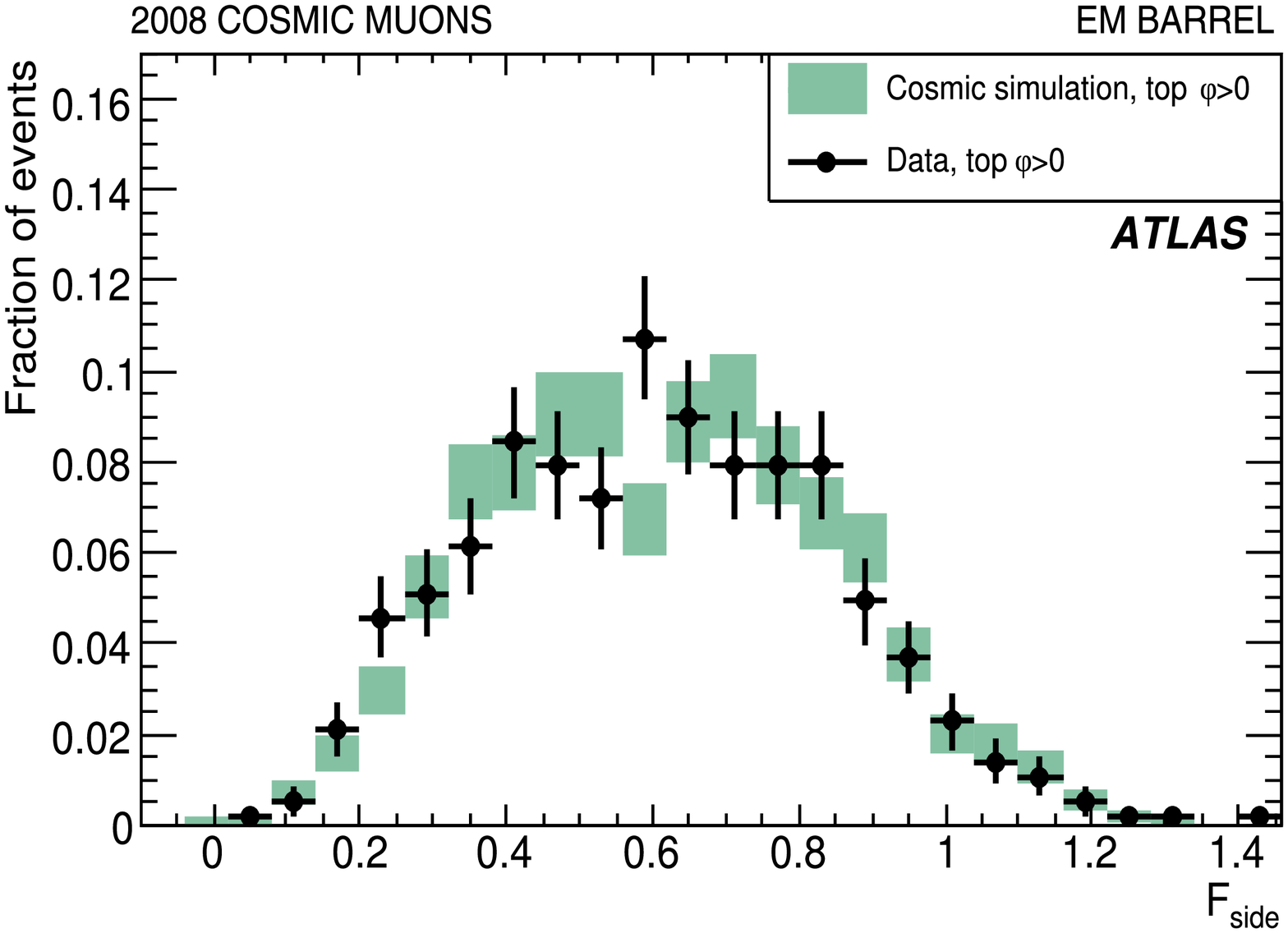}
\caption{ \em Lateral shower containment in the first layer for
  ``collision-like'' (top panel) or ``reverse''
  (bottom panel)  
  electromagnetic showers for radiative cosmic muon data
  (dots) and Monte Carlo simulation (rectangles). The definition of
  the $F_{\rm side}$ is given in the text. } 
\label{fig:photons2}
\end{center}
\end{figure}

Various shower shape distributions used for photon identification have been
compared with the Monte Carlo simulation: Figures~\ref{fig:photons1b}
and~\ref{fig:photons2} show two distributions of
variables related to lateral shower containment in the first
and second layers of the electromagnetic calorimeter. 

Figure~\ref{fig:photons1b} shows the ratio of the energy
deposited in a $\Delta \eta \times \Delta \phi = 3\times 7$ (in second
layer cell unit) cluster to
that in a $7\times 7$ cluster, in the 
second layer of the barrel calorimeter. In LHC
collisions, this variable
distinguishes electromagnetic showers, contained in $3$ cells in
$\eta$, from hadronic showers, leaking outside these $3$ cells. 
The contribution from the noise 
explains that the ratio can be above 1.

Figure~\ref{fig:photons2} shows the variable 
$F_{\rm side}=(E_{\pm3}-E_{\pm 1})/E_{\pm 1}$ computed as the ratio 
of energy within seven central cells in the first 
layer ($E_{\pm 3}$), outside a core of three central cells ($E_{\pm 1}$), 
over energy in the three central cells : in LHC collisions, this variable typically
separates photons, where little energy is deposited outside
the core region, from $\pi^0$s, where the two photons produced by the
$\pi^0$ deposit some energy outside the core region. 
The agreement between
the Monte Carlo simulation and the cosmic ray data is very good in both the cases
where the electromagnetic shower develops in the ``collision-like''
direction (in the bottom hemisphere) and the case where it develops in the backward 
direction (in the top hemisphere). 

Within the statistics available from data, important calorimeter
variables used in the electron/photon identification in ATLAS illustrate the
good agreement between the Monte Carlo simulation and electromagnetic
showers from radiative cosmic events in the calorimeter. These results, 
as well as the numerous comparisons done with beam test
data~\cite{ATLAS_TB_EM1,ATLAS_TB_EM2,ATLAS_TB_EM3,ATLAS_TB_EM4,ATLAS_TB_EM5}, give
confidence that robust photon and electron identification will be available for
early data at the LHC.

%% file: Conclusion.tex
\section{Conclusions and Perspectives}
\label{sec:conclu}

The liquid argon calorimeter has been installed, connected  and
fully readout since the beginning of 2008. Since then, much experience has been
gained in operating the system. Thanks to the very stable cryogenics and 
electronics operation over this period, first performance studies with the
complete LAr calorimeter coverage have been done using several months of
cosmic muon data and with LHC beam splash events from September 2008. These data
provided a check of the first level trigger energy 
computation and the timing of the electronics. 
In the EM calorimeter, detailed studies of the signal shape
predictions allow to check that, within the accuracy of the analysis,
there is no extra contribution to the dominant contributions to the
intrinsic constant term of the energy resolution. This indicates that
the reach of a global constant term of $0.7\%$ is achievable.  
The non-uniformity 
of the EM barrel calorimeter response to cosmic muons is consistent at the percent
level with the simulated response. Finally, the electromagnetic shower profiles
are in good agreement with the simulated ones, thus validating the
Monte Carlo description.  All these results allow for strong confidence  
in the readiness of the LAr calorimeter for the first LHC collisions. 

The ultimate LAr calorimeter
performance will be assessed with collision data: this is
particularly true for the electromagnetic and hadronic energy scale
computation in the ATLAS environment, which is needed for many ATLAS physics
analyses.







%% file: Acknowledgement-25may09.tex
\section{Acknowledgements}

We are greatly indebted to all CERN's departments and to the LHC project for their immense efforts not only in building the LHC, but also for their direct contributions to the construction and installation of the ATLAS detector and its infrastructure. We acknowledge equally warmly all our technical colleagues in the collaborating institutions without whom the ATLAS detector could not have been built. Furthermore we are grateful to all the funding agencies which supported generously the construction and the commissioning of the ATLAS detector and also provided the computing infrastructure.

The ATLAS detector design and construction has taken about fifteen years, and our thoughts are with all our colleagues who sadly could not see its final realisation.

We acknowledge the support of ANPCyT, Argentina; Yerevan Physics Institute, Armenia; ARC and DEST, Australia; Bundesministerium f\"ur Wissenschaft und Forschung, Austria; National Academy of Sciences of Azerbaijan; State Committee on Science \& Technologies of the Republic of Belarus; CNPq and FINEP, Brazil; NSERC, NRC, and CFI, Canada; CERN; NSFC, China; Ministry of Education, Youth and Sports of the Czech Republic, Ministry of Industry and Trade of the Czech Republic, and Committee for Collaboration of the Czech Republic with CERN; Danish Natural Science Research Council; European Commission, through the ARTEMIS Research Training Network; IN2P3-CNRS and Dapnia-CEA, France; Georgian Academy of Sciences; BMBF, DESY, DFG and MPG, Germany; Ministry of Education and Religion, through the EPEAEK program PYTHAGORAS II and GSRT, Greece; ISF, MINERVA, GIF, DIP, and Benoziyo Center, Israel; INFN, Italy; MEXT, Japan; CNRST, Morocco; FOM and NWO, Netherlands; The Research Council of Norway; Ministry of Science and Higher Education, Poland; GRICES and FCT, Portugal; Ministry of Education and Research, Romania; Ministry of Education and Science of the Russian Federation, Russian Federal Agency of Science and Innovations, and Russian Federal Agency of Atomic Energy; JINR; Ministry of Science, Serbia; Department of International Science and Technology Cooperation, Ministry of Education of the Slovak Republic; Slovenian Research Agency, Ministry of Higher Education, Science and Technology, Slovenia; Ministerio de Educaci\'{o}n y Ciencia, Spain; The Swedish Research Council, The Knut and Alice Wallenberg Foundation, Sweden; State Secretariat for Education and Science, Swiss National Science Foundation, and Cantons of Bern and Geneva, Switzerland; National Science Council, Taiwan; TAEK, Turkey; The Science and Technology Facilities Council and The Leverhulme Trust, United Kingdom; DOE and NSF, United States of America.

%% file: Biblio.tex
{}